\documentclass[longauth]{aa}  
\usepackage{graphicx}
\usepackage{txfonts}
\usepackage{xcolor}
\usepackage{tabularx}
\usepackage{booktabs}
\usepackage{orcidlink}
\newcommand{\collaboration}[1]{\unskip~#1}
\begin{document}

\title{Gaia Science Alerts microlensing events with mass estimates verifiable in Gaia Data Release 4}
\authorrunning{Wicker et al.}
\titlerunning{Gaia Science Alerts microlensing events in Gaia DR4}

\author{
    M. Wicker\inst{\ref{aff:UW},\ref{aff:ING}}\orcidlink{0009-0006-7642-0943}
    \and
    {\L}. Wyrzykowski\inst{\ref{aff:UW}, \ref{aff:NCBJ}}\orcidlink{0000-0002-9658-6151}
    \and
    M. Hundertmark\inst{\ref{aff:ARI}}\orcidlink{0000-0003-0961-5231}
    \and
    K. Kruszy\'{n}ska\inst{\ref{aff:LCO}}\orcidlink{0000-0002-2729-5369}
    \and
    E. Bachelet\inst{\ref{aff:CNRSBesancon}}
    \and
    P. Zieli\'nski\inst{\ref{aff:Torun}}\orcidlink{0000-0001-6434-9429}
    \and
    M. Kalicki\inst{\ref{aff:Torun}}
    \and
    K. Kotysz\inst{\ref{aff:UW},\ref{aff:Wroclaw}}\orcidlink{0000-0003-4960-7463}
    \and
    P. J. Miko{\l}ajczyk\inst{\ref{aff:NCBJ},\ref{aff:Wroclaw}}\orcidlink{0000-0001-8916-8050}
    \and
    M. Gromadzki\inst{\ref{aff:UW}}\orcidlink{0000-0002-1650-1518}
    \and
    K. Howil\inst{\ref{aff:UW}}
    \and
    J. Majumdar\inst{\ref{aff:UW}}\orcidlink{0009-0008-3960-1213}
    \and
    N. Ihanec\inst{\ref{aff:ING}}\orcidlink{0009-0003-7672-4806}
    \and
    M. Jab{\l}o\'{n}ska\inst{\ref{aff:UW}, \ref{aff:AustraliaStromlo}}
    \and
    M. Ratajczak\inst{\ref{aff:UW}}\orcidlink{0000-0002-3218-2684}
    \and
    M. Sitek\inst{\ref{aff:UW}}
    \and
    E. Stonkut\.{e}\inst{\ref{aff:Vilnius}}\orcidlink{0000-0002-8028-8133}
    \and
    M. Maskoli\=unas\inst{\ref{aff:Vilnius}}
    \and
    E. Pak\v{s}tien\.{e}\inst{\ref{aff:Vilnius}}\orcidlink{0000-0002-3326-2918}
    \and
    J. Zdanavi\v{c}ius\inst{\ref{aff:Vilnius}}
    \and
    V. \v{C}epas\inst{\ref{aff:Vilnius}}
    \and
    U. Jonauskait\.{e}\inst{\ref{aff:Vilnius}}\orcidlink{0009-0000-0293-8834}
    \and \\ 
    V. Bozza\inst{\ref{aff:Salerno},\ref{aff:Napoli}}
    \and
    A. Cassan\inst{\ref{aff:Paris}}
    \and
    M. Dominik\inst{\ref{aff:StAndrew}}
    \and
    R. Figuera Jaimes\inst{\ref{aff:MASchile},\ref{aff:PUCchile}}
    \and
    M. Rabus\inst{\ref{aff:ConcepcionChile}}\orcidlink{0000-0003-2935-7196}
    \and
    P. Rota\inst{\ref{aff:Salerno},\ref{aff:Napoli}}\orcidlink{0000-0001-6161-1604}
    \and
    K. A. Rybicki\inst{\ref{aff:UW}}\orcidlink{0000-0002-9326-9329}
    \and
    R. A. Street\inst{\ref{aff:LCO}}
    \and
    Y. Tsapras\inst{\ref{aff:ARI}}\orcidlink{0000-0001-8411-351X}
    \and
    J. Wambsganss\inst{\ref{aff:ARI}}
    \collaboration{(The OMEGA Key Project)} \and \\
    S. Awiphan\inst{\ref{aff:Thailand}}\orcidlink{0000-0003-3251-3583}
    \and
    T. Butterley\inst{\ref{aff:uniDurham}}\orcidlink{0000-0002-2853-0834}
    \and
    U. Burgaz\inst{\ref{aff:dublin}}\orcidlink{0000-0003-0126-3999}
    \and
    J.~M.~Carrasco\inst{\ref{aff:barcelonaUB},\ref{aff:cienciesICCUB},\ref{aff:catalunyaIEEC}}\orcidlink{0000-0002-3029-5853}
    \and
    F. Cusano\inst{\ref{aff:inafBologna}}
    \and
    V. S. Dhillon\inst{\ref{aff:uniSheffield},\ref{aff:iacTenerife}}
    \and
    R. Wilson\inst{\ref{aff:uniDurham}}
    \and
    H. H. Esenoglu\inst{\ref{aff:istanbul}}\orcidlink{0000-0003-3531-7510}
    \and
    A. Garofalo\inst{\ref{aff:inafBologna}}
    \and
    M. Jelinek\inst{\ref{aff:czechrepublic}}
    \and
    F. Lewis\inst{\ref{aff:faulkesTelescope},\ref{aff:Liverpool}}\orcidlink{0000-0003-3352-2334}
    \and
    U. Pylypenko\inst{\ref{aff:Torku},\ref{aff:finnishFINCA}}\orcidlink{0009-0002-7560-1903}
    \and
    J.-P. Rivet\inst{\ref{aff:coteazure}}\orcidlink{0000-0002-0289-5851}
    \and
    O. Tasuya\inst{\ref{aff:Thailand}}
    \and
    P. Trzcionkowski\inst{\ref{aff:UW}} \,
    \collaboration{(BHTOM Team)} \and \\
    F. Abe\inst{\ref{aff:nagoyaSpaceEarth}}
    \and
    D.~P.~Bennett\inst{\ref{aff:nasagoddard},\ref{aff:uniMaryland}}
    \and
    Aparna Bhattacharya\inst{\ref{aff:nasagoddard},\ref{aff:uniMaryland}}
    \and
    I. A. Bond\inst{\ref{aff:masseyUni}}
    \and
    K. Daimon\inst{\ref{aff:osakaSpaceEarth}}
    \and
    R. Hamada\inst{\ref{aff:osakaSpaceEarth}}
    \and
    Y. Hirao\inst{\ref{aff:unitokio}}
    \and
    S. Makida\inst{\ref{aff:osakaSpaceEarth}}
    \and
    S. Ishitani Silva\inst{\ref{aff:catholicaWashington},\ref{aff:nasagoddard}}
    \and
    S. Miyazaki\inst{\ref{aff:japanaerospace}}
    \and
    Y. Muraki\inst{\ref{aff:nagoyaSpaceEarth}}
    \and
    T. Nagai\inst{\ref{aff:osakaSpaceEarth}}
    \and
    K. Nunota\inst{\ref{aff:osakaSpaceEarth}}
    \and
    R. Ogawa\inst{\ref{aff:osakaSpaceEarth}}
    \and
    R. Oishi\inst{\ref{aff:osakaSpaceEarth}}
    \and
    H. Ose\inst{\ref{aff:osakaSpaceEarth}}
    \and
    G. Olmschenk\inst{\ref{aff:nasagoddard}}
    \and
    C. Ranc\inst{\ref{aff:Paris}}
    \and
    N. J. Rattenbury\inst{\ref{aff:auckland}}
    \and
    Y. K. Satoh\inst{\ref{aff:kantogakuinJapan}}
    \and
    T. Sumi\inst{\ref{aff:osakaSpaceEarth}}
    \and
    D. Suzuki\inst{\ref{aff:osakaSpaceEarth}}
    \and
    T. Tamaoki\inst{\ref{aff:osakaSpaceEarth}}
    \and
    S. K. Terry\inst{\ref{aff:nasagoddard},\ref{aff:uniMaryland}}
    \and
    C. Ueda\inst{\ref{aff:osakaSpaceEarth}}
    \and
    P. J. Tristram\inst{\ref{aff:canterburyNZ}}
    \and
    A. Vandorou\inst{\ref{aff:nasagoddard},\ref{aff:uniMaryland}}
    \and
    H. Yama\inst{\ref{aff:osakaSpaceEarth}} \,
    \collaboration{(MOA collaboration)}
}
\institute{
    \label{aff:UW}Astronomical Observatory, University of Warsaw, Al. Ujazdowskie 4, 00-478 Warsaw, Poland \\
    \email{mwicker@astrouw.edu.pl}
    \and
    \label{aff:ING}Isaac Newton Group of Telescopes, Apartado 321, E-38700 Santa Cruz de La Palma, Spain
    \and
    \label{aff:NCBJ}Astrophysics Division, National Centre for Nuclear Research, Pasteura 7, 02-093, Warsaw, Poland
    \and
    \label{aff:ARI}Astronomisches Rechen-Institut, Zentrum f{\"u}r Astronomie der Universit{\"a}t Heidelberg (ZAH), 69120 Heidelberg, Germany
    \and
    \label{aff:LCO}Las Cumbres Observatory, 6740 Cortona Drive, Suite 102, 93117 Goleta, CA, USA
    \and
    \label{aff:CNRSBesancon}Universit{\'e} Marie et Louis Pasteur, CNRS, Institut UTINAM UMR 6213, Besan\c{c}on, France
    \and
    \label{aff:Torun}Institute of Astronomy, Faculty of Physics, Astronomy and Informatics, Nicolaus Copernicus University, Grudzi\k{a}dzka 5, 87-100 Toru{\'n}, Poland
    \and
    \label{aff:Wroclaw}Astronomical Institute, University of Wroc{\l}aw, ul. Miko{\l}aja Kopernika 11, 51-622 Wroc{\l}aw, Poland
    \and
    \label{aff:AustraliaStromlo}Research School of Astronomy and Astrophysics, Australian National University, Mount Stromlo Observatory, Cotter Road Weston Creek, ACT 2611, Australia
    \and
    \label{aff:Vilnius}Vilnius University, Faculty of Physics, Institute of Theoretical Physics and Astronomy, Sauletekio av. 3, 10257 Vilnius, Lithuania
    \and
    \label{aff:Salerno}Dipartimento di Fisica "E.R. Caianiello", Universit{\`a} di Salerno, Via Giovanni Paolo II 132, I-84084 Fisciano, Italy
    \and
    \label{aff:Napoli}Istituto Nazionale di Fisica Nucleare, Sezione di Napoli, Via Cintia, I-80126 Napoli, Italy
    \and
    \label{aff:Paris}Institut d’Astrophysique de Paris, Sorbonne Universit{\'e}, CNRS, UMR 7095, 98 bis bd Arago, F-75014 Paris, France
    \and
    \label{aff:StAndrew}University of St Andrews, School of Physics \& Astronomy, St Andrews Centre for Exoplanet Science (StACES), North Haugh, St Andrews, KY16 9SS, United Kingdom
    \and
    \label{aff:MASchile}Millennium Institute of Astrophysics MAS, Nuncio Monsenor Sotero Sanz 100, Of. 104, Providencia, Santiago, Chile
    \and
    \label{aff:PUCchile}Instituto de Astrof{\'i}sica, Facultad de F{\'i}sica, Pontificia Universidad Cat{\'o}lica de Chile, Av. Vicu{\~n}a Mackenna 4860, 7820436 Macul, Santiago, Chile
    \and
    \label{aff:ConcepcionChile}Departamento de Matem{\'a}tica y F{\'i}sica Aplicadas, Facultad de Ingenier{\'i}a, Universidad Cat{\'o}lica de la Sant{\'i}sima Concepci{\'o}n, Alonso de Rivera 2850, Concepci{\'o}n, Chile
    \and
    \label{aff:Thailand}National Astronomical Research Institute of Thailand (Public Organization), 260 Moo 4, Donkaew, Mae Rim, Chiang Mai 50180, Thailand
    \and
    \label{aff:uniDurham}Centre for Advanced Instrumentation, Department of Physics, Durham University, South Road, DH1 3LE, UK    
    \and
    \label{aff:dublin}School of Physics, Trinity College Dublin, The University of Dublin, Dublin 2, Ireland
    \and
    \label{aff:barcelonaUB}Departament de F\'{\i}sica Qu\`antica i Astrof\'{\i}sica (FQA), Universitat de Barcelona (UB), Mart\'{\i} i Franqu\`es 1, E-08028 Barcelona, Spain
    \and
    \label{aff:cienciesICCUB}Institut de Ci\`encies del Cosmos (ICCUB), Universitat de Barcelona (UB), Mart\'{\i} i Franqu\`es 1, E-08028 Barcelona, Spain
    \and
    \label{aff:catalunyaIEEC}Institut d'Estudis Espacials de Catalunya (IEEC), Edifici RDIT, Campus UPC, 08860 Castelldefels (Barcelona), Spain
    \and
    \label{aff:inafBologna}INAF Osservatorio di Astrofisica e Scienza dello Spazio di Bologna, via Piero Gobetti 93/3, 40129 Bologna, Italy
    \and
    \label{aff:uniSheffield}Astrophysics Research Cluster, School of Mathematical and Physical Sciences, University of Sheffield, Sheffield S3 7RH, UK
    \and
    \label{aff:iacTenerife}Instituto de Astrof\'isica de Canarias, E-38205 La Laguna, Tenerife, Spain
    \and
    \label{aff:istanbul}Istanbul University, Faculty of Science, Department of Astronomy and Space Sciences, 34119, Istanbul, Türkiye
    \and
    \label{aff:czechrepublic}Astronomical Institute of the Academy of Sciences of the Czech Republic (ASU CAS), 25165 Ondrejov, CZ
    \and
    \label{aff:faulkesTelescope}Faulkes Telescope Project, Cardiff, Wales, UK
    \and
    \label{aff:Liverpool}The Schools’ Observatory, Astrophysics Research Institute, Liverpool John Moores University, 146 Brownlow Hill, Liverpool, UK
    \and
    \label{aff:Torku}Department of Physics and Astronomy, University of Turku, FI-20014 Turku, Finland
    \and
    \label{aff:finnishFINCA}Finnish Centre for Astronomy with ESO (FINCA), FI-20014 University of Turku, Finland
    \and
    \label{aff:coteazure}Universit{\'e} C{\^o}te d’Azur, CNRS, OCA, LAGRANGE, France
    \and
    \label{aff:nagoyaSpaceEarth}Institute for Space-Earth Environmental Research, Nagoya University, Nagoya 464-8601, Japan
    \and
    \label{aff:nasagoddard}Code 667, NASA Goddard Space Flight Center, Greenbelt, MD 20771, USA
    \and
    \label{aff:uniMaryland}Department of Astronomy, University of Maryland, College Park, MD 20742, USA
    \and
    \label{aff:masseyUni}School of Mathematical and Computational Sciences, Massey University, Auckland 0745, New Zealand
    \and
    \label{aff:osakaSpaceEarth}Department of Earth and Space Science, Graduate School of Science, The University of Osaka, Toyonaka, Osaka 560-0043, Japan
    \and
    \label{aff:unitokio}Institute of Astronomy, Graduate School of Science, The University of Tokyo, 2-21-1 Osawa, Mitaka, Tokyo 181-0015, Japan
    \and
    \label{aff:catholicaWashington}Department of Physics, The Catholic University of America, Washington, DC 20064, USA
    \and
    \label{aff:japanaerospace}Institute of Space and Astronautical Science, Japan Aerospace Exploration Agency, 3-1-1 Yoshinodai, Chuo, Sagamihara, Kanagawa 252-5210, Japan
    \and
    \label{aff:auckland}Department of Physics, University of Auckland, Private Bag 92019, Auckland, New Zealand
    \and
    \label{aff:kantogakuinJapan}College of Science and Engineering, Kanto Gakuin University, Yokohama, Kanagawa 236-8501, Japan
    \and
    \label{aff:canterburyNZ}University of Canterbury Mt. John Observatory, P.O. Box 56, Lake Tekapo 8770, New Zealand
}

\date{Received 21 August 2026; accepted 16 October 2026}

\abstract{
{Gaia Science Alerts provide an all-sky stream of photometric transients, including events consistent with gravitational microlensing. For such events, forthcoming epoch astrometry from the Gaia mission offers the possibility of testing photometric lens mass predictions. While microlensing parallax inferred from the photometric light curve can constrain the event geometry, a direct lens mass measurement additionally requires the angular Einstein radius, which may be accessible through astrometric microlensing.}
{We aim to construct a pre-Data-Release-4 (DR4) sample of Gaia microlensing events for which the available photometry provides a measurable microlensing-parallax signal. This allows us to derive statistical predictions for the lens mass and distance. We forward-modelled these events to predict the detectability of the astrometric microlensing signal in Gaia DR4.}
{We identify 56 Gaia Science Alerts microlensing events with 71 point-source point-lens solutions with annual parallaxes that satisfy our fit-quality, posterior-boundedness, and parallax-significance criteria. Lens masses and distances were inferred using two prior-based frameworks, pyLIMASS and the Besançon Galaxy Model, which combine the measured microlensing parameters and Gaia source information with stellar-population and Galactic models. These inferred parameters were used to simulate Gaia DR4-like along-scan astrometry and estimate the detectability of the astrometric microlensing signal. The inferred lens masses and distances remain strongly prior-dependent without a direct angular Einstein radius measurement.}
{Among the 284 event--solution--prior branches, 47 are classified as high-confidence astrometric detections using Gaia DR4-like simulations. These predictions can be tested once epoch astrometry is released in Gaia DR4.}
{}
}

\maketitle

\section{Introduction}

Gravitational microlensing occurs when a foreground lens passes sufficiently close to the line of sight of a background source, producing transient magnification  \citep{Paczynski1986,Gould2000}. Because the signal is gravitational rather than radiative, microlensing is sensitive to lenses regardless of their luminosity and, therefore, provides a unique probe of faint or non-luminous objects. This includes low-mass stars, isolated stellar remnants \citep{SahuBH}, planets, and free-floating planetary-mass objects.

Since the first detections in the early 1990s, microlensing has been driven primarily by photometric monitoring: first by surveys such as MACHO, EROS, and OGLE \citep{Alcock1993,Aubourg1993,Udalski1993}, and later by dedicated high-cadence Galactic-bulge surveys such as OGLE, MOA, KMTNet, and PRIME \citep{Udalski2015,Bond2001,Kim2016,SumiPrime}. This is complemented by wide-field transient surveys such as ASAS-SN, ATLAS, Gaia, VVV, and ZTF \citep{Shappee2014,Tonry2018,GaiaCollaboration2016,MinnitiVVV,BellmZTF}. In the coming decade, the Vera C. Rubin Observatory's Legacy Survey of Space and Time (LSST) and the Nancy Grace Roman Space Telescope will further transform the field. The Rubin observatory will provide wide-field optical monitoring across much of the southern sky, while the Roman telescope will conduct high-precision near-infrared time-domain observations through the Roman Galactic Bulge Time Domain Survey and the Roman Galactic Plane Survey, delivering large new samples of microlensing events \citep{IvezicLSST,RomanPenny}.

The physical scale and duration of a microlensing event are determined by three fundamental quantities: the lens mass ($M_{\rm L}$), the lens--source relative parallax ($\pi_{\rm rel}$), and the lens--source relative proper-motion vector ($\boldsymbol{\mu}_{\rm rel}$). These determine the angular Einstein radius ($\theta_{\rm E}=(\kappa M_{\rm L}\pi_{\rm rel})^{1/2}$), and the Einstein timescale ($t_{\rm E}=\theta_{\rm E}/\mu_{\rm rel}$), where $\mu_{\rm rel}=|\boldsymbol{\mu}_{\rm rel}|$ and $\kappa=4G/(c^2\,{\rm au})$. The microlensing-parallax vector ($\boldsymbol{\pi}_{\rm E}$) has magnitude $\pi_{\rm E}=\pi_{\rm rel}/\theta_{\rm E}$ and points in the direction of $\boldsymbol{\mu}_{\rm rel}$. Photometric light curves generally constrain $t_{\rm E}$ and, when higher-order distortions are detected, $\boldsymbol{\pi}_{\rm E}$, but these quantities alone do not provide a direct measurement of the lens mass or distance.

Additional constraints can be obtained from annual or satellite microlensing parallaxes, finite-source effects, lens-flux constraints, spectroscopic characterisation of the source star, high-resolution imaging, interferometric observations, and astrometric microlensing. In particular, combining the microlensing-parallax amplitude with the angular Einstein radius yields the lens mass directly through $M_{\rm L}=\theta_{\rm E}/(\kappa\pi_{\rm E})$ \citep{Gould2000}. Astrometric microlensing produces a displacement of the centroid of light of the combined unresolved lensed images relative to the unlensed source position. Because the amplitude of this displacement is proportional to $\theta_{\rm E}$, modelling the astrometric trajectory provides a measurement of the angular Einstein radius \citep{DominikiSahu2000}. Combining photometric and astrometric microlensing measurements can therefore constrain the lens mass and distance and break degeneracies inherent in photometric-only models.

Gaia is uniquely suited to studying microlensing events. The mission combines homogeneous all-sky photometry with high-precision astrometry and spectroscopic information for over two billion sources \citep{GaiaCollaboration2016}. The Gaia Science Alerts (GSA) system has identified and published ongoing photometric transients in near real time, enabling ground-based follow-up while transients continue to evolve \citep{Hodgkin2021}. For microlensing, Gaia is particularly complementary to ground-based bulge surveys because it covers the Galactic disc and higher-latitude fields, where events can be bright, appear relatively nearby, and favour astrometric lensing \citep{Belokurov2002}. Gaia Data Release 3 (DR3) has already delivered the first all-sky catalogue of Gaia microlensing candidates \citep{Wyrzykowski2023} with multiple dark lens candidates \citep{Kasia2024}. 

Gaia photometry has contributed to a diverse set of microlensing discoveries, including planetary events such as Gaia22dkvLb \citep{Wu2024}, AT2021ueyLb/Gaia21dnc \citep{ban2025}, and Gaia23bra b \citep{Harris_2026}, long-timescale dark-lens candidates such as Gaia18cbf \citep{Kruszynska18cbf} and Gaia18ajz \citep{Howil2025}, high-magnification events with a complementary space-parallax signal such as Gaia19bld \citep{Rybicki19bld}, and binary-lens events with full Keplerian orbital solutions such as Gaia16aye \citep{Wyrzykowski2020} and Gaia20fnr \citep{Wicker2026}. Gaia data have also contributed to the identification of very short-timescale events associated with free-floating or wide-orbit planetary-mass lenses, such as KMT-2024-BLG-0792/OGLE-2024-BLG-0516 \citep{Dong2026}. For Gaia DR3-ULENS-001, the anomalous Gaia DR3 astrometric parameters are consistent with the astrometric microlensing signal predicted by simulations based on the photometric model. This agreement suggests a nearby isolated white dwarf as the lens \citep{Jablonska2022}. Many more such events are expected in Gaia Data Release 4 (DR4)\footnote{Expected 2 December 2026}, where epoch astrometric and photometric data will be available for all sources. The long Gaia astrometric time series, paired with photometric data from other surveys or follow-up observations, makes Gaia sensitive to long-timescale microlensing events, such as those from isolated black holes \citep{RybickiMLGaiaOGLE}.

In this study we constructed a pre-DR4 benchmark sample of microlensing events discovered through GSA. We considered alerts issued before the Gaia DR4 time cut, identified events consistent with microlensing after automated filtering, cross-matching, modelling, and visual vetting, and focused on the subset for which the available photometry supports a point-source point-lens (PSPL) model with annual parallax.  We fitted the photometric models to the combined Gaia and follow-up data and separately to the Gaia-only light curves to quantify the role of complementary monitoring. We then inferred lens mass and distance posteriors using both a Galactic-prior and a Gaussian-mixture approach. Using these inferred lens properties, we forward-modelled the astrometric microlensing signals expected in Gaia DR4. Our goal is to provide testable pre-DR4 mass predictions and identify the Gaia-alert events for which future Gaia astrometry is most likely to verify the microlensing scenario.

This paper is organised as follows. In Sect.~\ref{sec:data} we describe the GSA parent sample, follow-up data, and event-selection procedure. In Sect.~\ref{sec:mlModel} we present the photometric model and the treatment of parallax solutions. In Sect.~\ref{sec:photometricSolutions} we summarise the sample of the final photometric models, which we compare to Gaia-only models in Sect.~\ref{sec:gaiaonlyvsfollowup}. In Sect.~\ref{sec:lensMassinference} we present the lens mass and distance estimates, and in Sect.~\ref{sec:predictAstrometry} we predict the expected Gaia DR4 astrometric microlensing signals. We discuss the implications and limitations of the sample and summarise our conclusions in Sect.~\ref{sec:conclusions}.

\section{Parent sample, data, and event selection}
\label{sec:data}
\begin{figure*}[t]
    \centering
    \includegraphics[width=\textwidth]{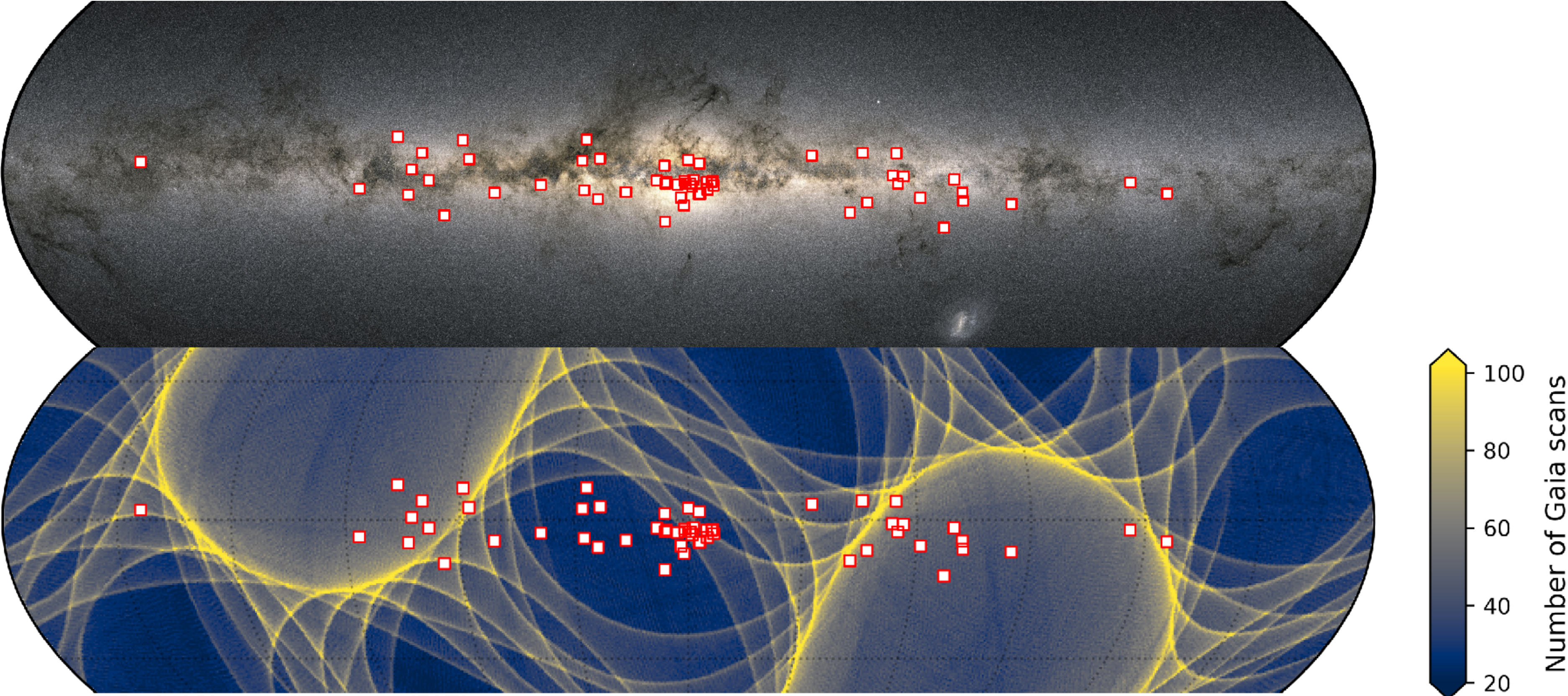}
    \caption{Sky map view of the Milky Way (top) and Gaia DR3 nominal scanning law map (bottom). The locations of the final sample of microlensing events are marked with red squares. The figures were made using \texttt{MW-Plot}, \emph{Gaia} EDR3 data \citep{GaiaEDR3}, and \texttt{gaiaunlimited}.}
    \label{fig:milkywayViewplot}
\end{figure*}

\subsection{Parent sample from Gaia Science Alerts}

The starting point for our catalogue of gravitational microlensing events for which epoch astrometric data will be released in Gaia DR4 was the full public archive of GSA. This archive provides a real-time transient stream generated from Gaia photometric observations \citep{Hodgkin2021}. The archive contains 27\,076 alerts covering the entire Gaia mission lifetime.
Each alert entry provides a uniform set of metadata, including the alert identifier, alert date, equatorial coordinates, alert magnitude, historical source brightness and scatter, assigned transient class, publication flag, comments, and transient name server (TNS) identifier when available. In addition, the alert pages provide the Gaia $G$-band light curves. Although per-epoch BP/RP spectra are displayed on the GSA webpages, these are not distributed in a calibrated form suitable for homogeneous use in the present analysis and were therefore not used further. All Gaia photometric models in this work are based on the calibrated Gaia $G$-band measurements released through the alert system.

The archive metadata are informative but were not adopted as hard selection criteria. There are 45 unique class labels in GSA. Among the 27\,076 alerts in the full archive, only 42 alerts carry the explicit \texttt{ULENS} class label. Likewise, the free-text comment field contains the string `ulens' or related lensing annotations for 715 alerts in the full archive. Importantly, both the class labels and comments reflect the information available at or around the time the alert was issued, when the transient light curve was often still sparsely sampled and its nature uncertain. Microlensing events can therefore be difficult to recognise reliably at this early stage, particularly before the characteristic light-curve morphology is fully developed. This illustrates the heterogeneous nature of real-time alert classification: potentially relevant microlensing events may carry the dedicated \texttt{ULENS} label, may only be mentioned in the comments, or may remain classified under another transient class. We therefore did not pre-select targets on the basis of the archive classification alone.

\subsection{Temporal cut and relevance to DR4}
According to the current public Gaia DR4 content summary\footnote{\url{https://www.cosmos.esa.int/web/gaia/data-release-4}}, DR4 is based on observations acquired between 25 July 2014 at 10:30 UTC and 20 January 2020 at 22:00 UTC, corresponding to a 66-month data interval.
As the present work is motivated by the prospect of complementing photometric microlensing models derived from GSA with astrometric information expected in Gaia DR4, the alert sample was restricted to events whose GSA alert was issued before the DR4 cut-off. This reduced the sample from 27\,076 to 11\,048 alerts.

Alerts issued after the DR4 cut-off were not considered, because their most informative microlensing epochs are unlikely to be represented in the DR4 astrometric time series. For long events, the main astrometric perturbation may occur after the final DR4 observation. For short events, the alert may occur too late for Gaia to have sampled the centroid shift sufficiently before the cut-off. By contrast, alerts issued before the cut-off offer the possibility that at least a substantial fraction of the relevant astrometric trajectory is represented within the DR4 time-series products. We emphasise that this choice introduces a time-dependent selection effect. The sample studied here is not intended to represent the full Gaia microlensing alert population. Rather, it represents the subset of alerts that pass the selection criteria described below and for which a pre-DR4 photometric model can, in principle, be tested against the forthcoming astrometric time-series data.

\subsection{Gaia time-series data}

For each focal-plane crossing, Gaia observes a source photometrically, astrometrically, and spectroscopically. The Gaia G-band alert light curves therefore provide the epochs at which Gaia astrometric information is also expected to be available in DR4. Even when a microlensing light curve can be well constrained by ground-based follow-up observations, the astrometric microlensing signal is measurable by Gaia only if it observed the source at a time near or during the magnified period. We therefore required a minimum number of Gaia G-band measurements in the alert light curve before considering an event for astrometric-signal predictions.

We retained only alerts with at least ten Gaia G-band measurements published through the GSA webpages. This cut also removes very faint events that were not detected by Gaia before the brightening that caused the alert. These events, Gaia17bzy and Gaia19dqr, therefore lack a Gaia DR3 source ID. The cut reduced the pre-DR4 alert sample from 11\,048 to 4\,839 candidate events for which we considered microlensing models. The sequential selection procedure of the full sample is listed in Table~\ref{tab:sample_selection}.

\begin{table*}
\caption{Sequential selection of the GSA microlensing sample.}
\label{tab:sample_selection}
\centering
\begin{tabular}{cllrr}
\hline\hline
\noalign{\vskip 1pt}
Step & Selection criterion & Sect. & Retained & Removed \\
\hline
\noalign{\vskip 3pt}
1 & Full GSA archive & 2.1 & 27\,076 & -- \\
2 & Alert issued before the DR4 cut-off (2020 January 20, 22:00 UTC) & 2.2 & 11\,048 & 16\,028 \\
3 & At least ten Gaia $G$-band measurements & 2.3 & 4\,839 & 6\,209 \\
4 & Unblended Gaia-only PSPL fit converges & 3.1 & 2\,665 & 2\,174 \\
5 & $\geq 2$ Gaia points within $t_0 \pm t_{\rm E}$ and $\geq 5$ outside & 3.1 & 513 & 2\,152 \\
6 & Cross-matching, external photometry, visual vetting & 3.2 & 229 & 284 \\
7 & Acceptance of PSPL models with parallax and remaining exclusions & 3.3--4 & 56 & 173 \\
\hline
\end{tabular}
\end{table*}

\section{Photometric modelling and final sample construction}
\label{sec:mlModel}
From our sample of GSA events, we identified candidate microlensing events by following a hierarchical procedure. We first used the Gaia alert photometry alone to identify light curves broadly compatible with a PSPL morphology. We then used catalogue cross-matches, external photometry, and visual inspection to remove clear contaminants and complex events. Finally, we performed homogeneous PSPL modelling with and without annual parallax on the vetted sample. This strategy follows the same general logic as previous large-scale Gaia microlensing searches \citep{Kasia2024}. All microlensing models were fitted with \texttt{pyLIMA}\footnote{\url{https://github.com/ebachelet/pyLIMA}} \citep{Bachelet2017}.

\subsection{Initial Gaia-only PSPL screening}
\label{sec}

We adopted the standard PSPL model.  For rectilinear lens--source motion, the instantaneous angular separation between the lens and source, in units of the angular
Einstein radius, is
\begin{equation}
    u(t) =
    \left[
    u_0^2 +
    \left(
    \frac{t-t_0}{t_{\rm E}}
    \right)^2
    \right]^{1/2},
\end{equation}
where $t_0$ is the time of closest approach, $u_0$ is the signed minimum lens--source separation in units of $\theta_{\rm E}$, and $t_{\rm E}$ is the Einstein-radius crossing time. The corresponding
PSPL magnification is

\begin{equation}
    A(u) = \frac{u^2+2} {u\sqrt{u^2+4}}.
\end{equation}

For each independent telescope and filter dataset $j$, the model flux is

\begin{equation}
    F_j(t) = f_{{\rm s},j} A[u(t)] + f_{{\rm b},j},
\end{equation}
where $f_{{\rm s},j}$ is the unmagnified source flux and $f_{{\rm b},j}$ is the unresolved blend flux. The flux coefficients were determined independently for each telescope and filter dataset.

For the initial screening, we used only the Gaia $G$-band light curves published through the GSA pages. We first fitted an unblended PSPL model with a Levenberg--Marquardt optimiser \citep{levenbergLM,marquardtLM}, using the parameters $(t_0,u_0,t_{\rm E},f_{\rm s,Gaia})$. The resulting solution was used only to initialise subsequent trust region reflective fits. We ran both blended and unblended optimisations at this stage but adopted the unblended solution for the automated screening because the blend term is poorly constrained for many sparsely sampled Gaia-only light curves and can otherwise drive unstable preliminary solutions. Blending was included again in all final modelling steps.

A GSA alert was retained as a preliminary microlensing candidate if the unblended Gaia-only PSPL fit converged to a finite solution. This reduced the pre-selected GSA alert sample from 4\,839 to 2\,665. We then required the fitted light curve to contain at least two Gaia measurements within $t_0\pm t_{\rm E}$ and at least five measurements outside this interval. These baseline measurements may lie on either side of the event. We did not require Gaia baseline coverage both before and after the magnified part of the light curve. This avoids removing long-timescale events and events whose magnified part overlaps with the beginning of the Gaia time series, for which baseline data are available on one side only. These cuts ensure that the Gaia data contain both event and baseline information, which is required for any later comparison with Gaia astrometry. These cuts also remove alerts for which a single bright point or a poorly sampled excursion cannot distinguish microlensing from other forms of variability. A visual inspection of the rejected long-timescale tail did not reveal convincing microlensing candidates lost by this criterion. The sample was thereby reduced from 2\,665 to 513 alerts.

We did not impose a strict $\chi^2$ threshold during this initial stage. A simple PSPL model can fit genuine events poorly when annual parallax, binary lens effects, binary source effects, or survey systematics are present. The purpose of the Gaia-only screening was therefore not to define the final sample but to select a manageable set of events for cross-matching, external-data collection, and detailed modelling.

\subsection{Cross-matching, external photometry, and data preparation}

The 513 preliminary candidates were cross-matched against external microlensing and transient surveys to identify previously recognised events and to assemble complementary photometry. We searched the alert and event lists of OGLE, KMTNet, MOA, ASAS-SN, ATLAS, and ZTF. This recovered counterparts to 64 OGLE, 69 KMTNet, 31 MOA, and 13 ZTF microlensing events within the Gaia DR4 time interval. These numbers are not mutually exclusive, as many events were detected by more than one survey.

We also cross-matched the candidates against catalogues of likely contaminants. These included young stellar objects (YSOs) from the Gaia--AllWISE and Konkoly Optical YSO catalogues (\citeauthor{Marton2019} \citeyear{Marton2019}, \citeyear{Marton2023}), active galactic nuclei (AGN) and quasars from the Milliquas catalogue \citep{Flesch2023}, Gaia DR3 variable AGN \citep{Carnerero2023}, nearby galaxies from GLADE+ \citep{Dalya2022}, and classified transients from the TNS \citep{GalYam2021}. These matches were used as vetting information rather than as automatic rejection criteria. Only clear non-microlensing contaminants were removed.

For events with available external data, we collected public photometry from KMTNet, MOA, and other survey pages \citep{Kim2016,Bond2001}. Although OGLE observations exist for several events in our sample, these data were not made available to us for this work. However, 53 of the 64 OGLE events were also monitored at high cadence by all three KMTNet telescopes, and therefore we do not expect the inclusion of the OGLE data to have significantly improved the constraints on the photometric light curves. These data were combined with follow-up observations available through the Black Hole Target and Observation Manager (BHTOM; \citealt{Wyrzykowski2024,Mikolajczyk2025}), an observation-coordinating and data-processing server built on the open-source TOM Toolkit developed at Las Cumbres Observatory \citep{Street2018,Street2024}. Each survey, observatory, and filter combination was treated as an independent dataset, so that the microlensing model could assign separate source and blend fluxes to each photometric system. In a small number of very sparse BHTOM follow-up datasets, observations from different small telescopes were combined.

Before final modelling, we applied uniform cleaning cuts to all datasets. We retained only measurements with finite magnitudes and uncertainties, $\sigma_m<0.2$ mag, and magnitudes in the range $0<m<22$. For each telescope and filter dataset separately, we calculated the median and standard deviation of the reported photometric uncertainties and removed measurements whose uncertainty exceeded the median by more than three standard deviations. For datasets containing more than 300 measurements, observations were grouped into one-day intervals. Within each bin, we calculated the inverse-variance weighted mean flux. The formal uncertainty of the binned flux was taken to be 
\begin{equation}
    \sigma_{\overline{F}} =
    \left(
    \sum_i \sigma_{F,i}^{-2}
    \right)^{-1/2}.
\end{equation}
This procedure assumes that measurements within each daily bin are statistically independent. No additional correlated-noise model was included.

After cross-matching, external-data collection, and visual inspection of the light curves, we retained 229 vetted microlensing candidates. These candidates formed the final sample for parallax modelling.

\subsection{Final PSPL modelling with and without annual parallax}
\label{sec}

For each vetted candidate we first fitted a blended PSPL model without parallax to the full multi-survey photometric dataset. We then repeated the fit with a blended PSPL model including annual microlensing parallax. Events whose light curves clearly required a different interpretation, such as binary-lens or binary-source models, or other non-PSPL behaviour, were excluded from the homogeneous sample.

The global optimisation was performed with the differential-evolution implementation in \texttt{pyLIMA} \citep{storn1997differential,2020SciPy-NMeth}. The parameter ranges were chosen to be broad enough to include the Gaia-alert population while avoiding unphysical solutions. Annual microlensing parallax was included using the geocentric formalism implemented in \texttt{pyLIMA}. All datasets, including the Gaia photometry, were modelled in this geocentric frame; the effect of Gaia's location at L2 is quantified in Sect.~4. The parallax model introduces the two components $\pi_{{\rm E},N}$ and $\pi_{{\rm E},E}$ of the microlensing-parallax vector. The event time $t_0$ was allowed to vary across the Gaia DR4 time interval $\pm 100$ days to avoid edge effects for long events. The impact parameter was restricted to $u_0\in[-2,2]$, the Einstein timescale to $t_{\rm E}\in[5,1000]$ d, and the parallax components to $\pi_{\rm E,N},\pi_{\rm E,E}\in[-1,1]$. 

Posterior exploration was then carried out with the \texttt{pyLIMA} Markov chain Monte Carlo (MCMC) implementation \citep{emceePaper} for both the blended PSPL and the combined blended PSPL and parallax models. The no-parallax solution provided the reference geometry and initial estimate of $t_0$ for the parallax fit. To explore the discrete degeneracies common in single-lens parallax modelling \citep{SmithparallaxDegen,GouldparallaxDegen}, we initialised the parallax chains from multiple sign combinations of $u_0$, $\pi_{\rm E,N}$, and $\pi_{\rm E,E}$. In many events these initialisations converged to the same posterior mode, while in others multiple separated solutions remained viable. Solutions were treated as distinct only when the recovered parameters differed by more than their $1\sigma$ uncertainties in at least one of the microlensing parameters. Pure sign realisations that were statistically indistinguishable were not counted as separate physical solutions.
Unless stated otherwise, the reported parameter estimates are the medians of the marginalised MCMC posteriors. The quoted $1\sigma$ uncertainties correspond to the 16th and 84th percentiles. For individual scalar parameters, a $3\sigma$ credible interval refers to the 0.135th to 99.865th percentile range.

\begin{figure}
    \centering
    \includegraphics[width=\columnwidth]{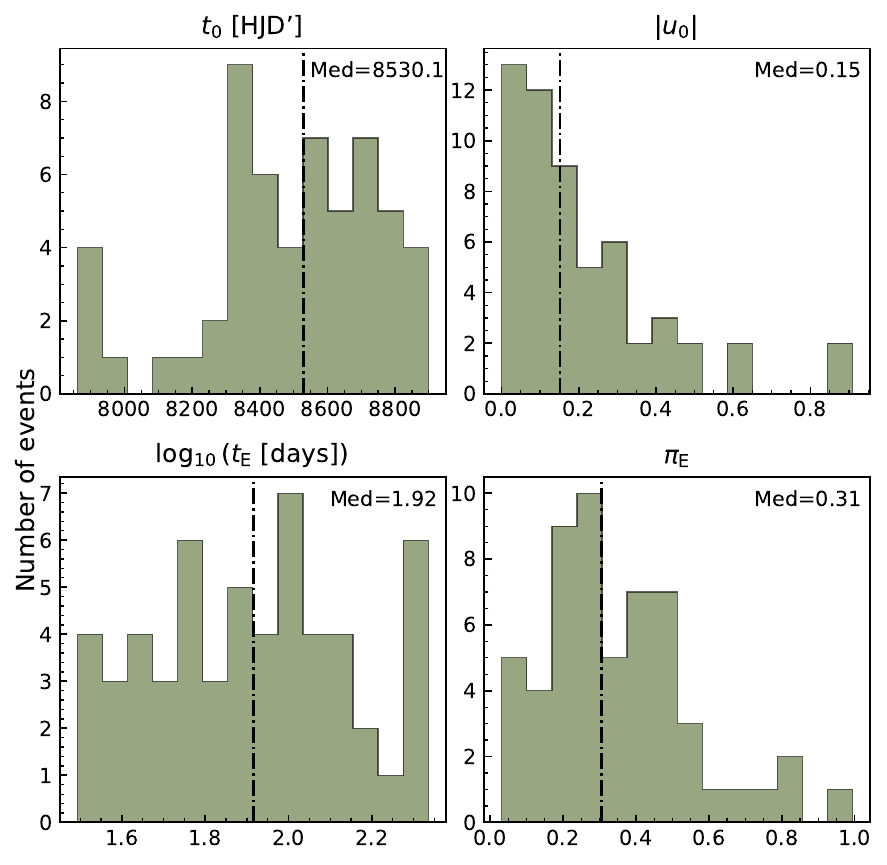}
    \caption{Distribution of the best-model microlensing parameters for the final sample of events. The vertical dash-dotted lines with the corresponding ‘Med’ labels mark the median values of the distributions.}
    \label{fig:microlensingParams_distrib}
\end{figure}

\begin{figure}
    \centering
    \includegraphics[width=\columnwidth]{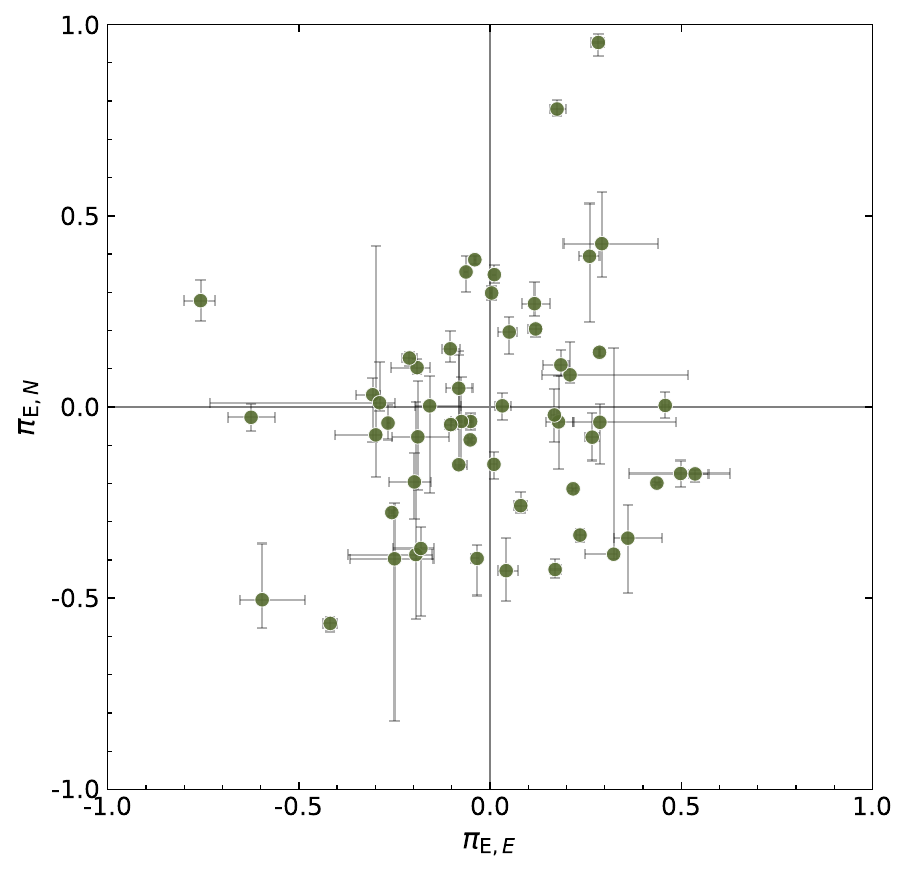}
    \caption{Distribution of the best-model parallax parameters for the final sample of events. The parallax components, $\pi_{\rm{E, N}}$ and $\pi_{\rm{E, E}}$, represent the northern and eastern components, respectively.}
    \label{fig:pllxParams_distrib}
\end{figure}

\begin{figure}
    \centering
    \includegraphics[width=\columnwidth]{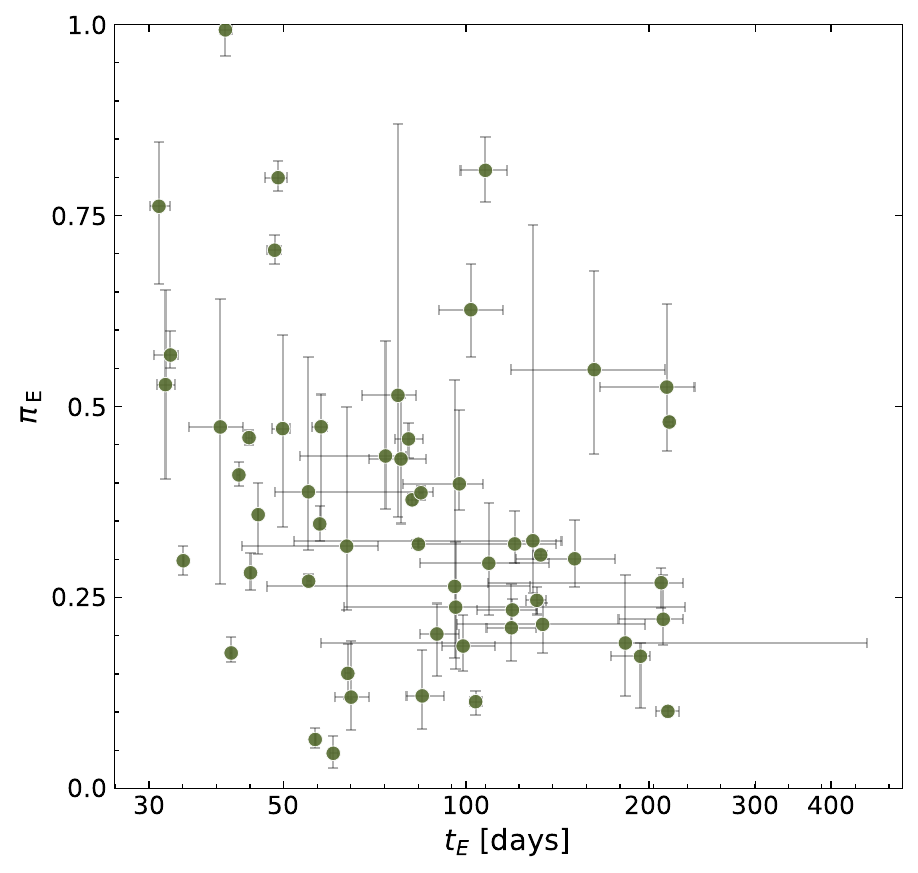}
    \caption{Distribution of the best-model $\pi_{\rm{E}}$ parameters as a function of the Einstein time ($t_{\rm{E}}$) for the final sample of events.}
    \label{fig:tepllxParams_distrib}
\end{figure}

\begin{figure}
    \centering
    \includegraphics[width=\columnwidth]{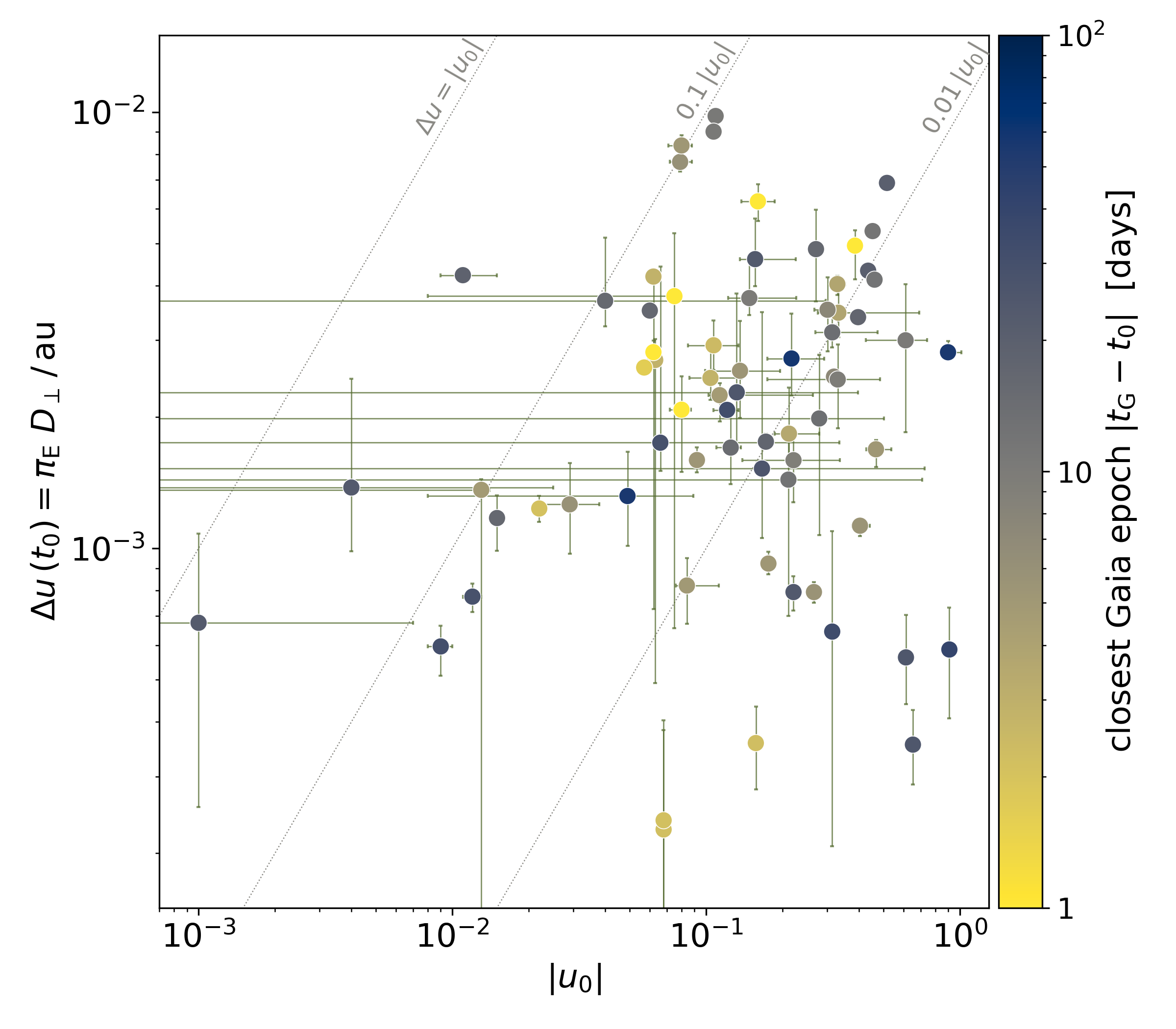}
    \caption{Difference between the lens--source separations seen from the Earth and from Gaia at L2, $\Delta u(t_0) = \pi_{\rm E} D_\perp/{\rm au}$, as a function of $|u_0|$ for the microlensing events in this study. $D_\perp$ is the projected Earth--Gaia separation at $t_0$. The diagonal dotted lines mark $\Delta u = |u_0|$, $0.1|u_0|$ and $0.01|u_0|$. The colour scale shows the time between $t_0$ and the closest Gaia epoch, $|t_G - t_0|$.}
    \label{fig:spacePllx_plot}
\end{figure}

\begin{figure*}[t]
    \centering
    \includegraphics[width=\textwidth]{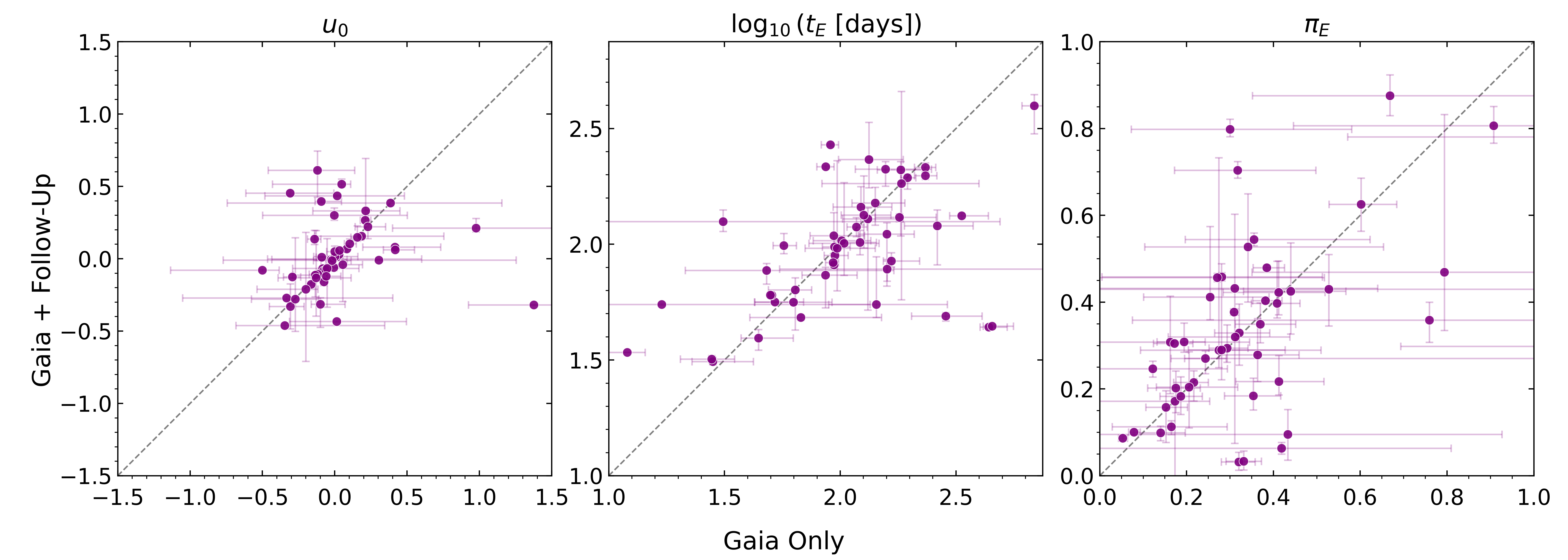}
    \caption{Comparison of the parameters recovered from Gaia-only refits and from the corresponding Gaia and follow-up solutions for the subset of Gaia-only runs that converge to physically acceptable solutions. The dashed lines indicate equality. While $u_0$ is often recovered approximately, the scatter increases for $t_{\rm E}$ and becomes most pronounced for the parallax amplitude ($\pi_{\rm E}$), illustrating the importance of complementary follow-up data for constraining higher-order microlensing parameters.}
    \label{fig:gaiaonly_compare}
\end{figure*}

\subsection{Acceptance of parallax solutions}

An event was retained in the final combined PSPL and parallax sample when the parallax solution satisfied three conditions. First, the parallax model had to provide a better fit than the corresponding no-parallax PSPL model. This was determined by comparing the $\chi^2$ of each solution. Second, the MCMC posterior had to be bounded and not cut off by the parameter ranges. Third, the parallax vector had to be inconsistent with $\pi_{\rm E}=0$ at the $3\sigma$ level. We did not automatically reject formally negative blend fluxes. In heterogeneous aperture and survey photometry, the fitted blend term can absorb calibration offsets, unresolved background structure, or baseline mismatches. This can lead to negative blend parameters. The solutions for which this was the case were inspected and retained only when the overall microlensing interpretation remained plausible.

We excluded from the homogeneous sample events that had already been studied in other work. These studies include Gaia16aye \citep{Wyrzykowski2020}, Gaia18ajz \citep{Howil2025}, Gaia18cbf \citep{Kruszynska18cbf}, Gaia19bld \citep{Rybicki19bld}, Gaia19dke \citep{Maskoliunas19dke}, and the events from \citet{RybickiBHevents} (Gaia18cej, Gaia18cho, Gaia19ayh, and Gaia19drv). Other events in the sample have been cross-matched with entries in other catalogues \citep[e.g.][]{Mroz2020, Medford2023}; however, because they have not been studied extensively on an individual basis, they were retained in the sample.

\section{Final photometric solutions}
\label{sec:photometricSolutions}

The final sample consists of 56 GSA microlensing events with 71 accepted PSPL and parallax solutions. By construction, this is not a complete sample of Gaia microlensing alerts. The sample is biased towards events that are sufficiently well covered photometrically, that are long enough for annual parallax to be measurable, and that are adequately described by a single-lens model. These selection effects are appropriate for the goal of this work: to identify Gaia-alerted microlensing events for which pre-DR4 photometric mass predictions can be tested with future Gaia astrometry.

The fitted parameters for these solutions are listed in Tables~\ref{tab:modelParamsGaia17},~\ref{tab:modelParamsGaia18}, and ~\ref{tab:modelParamsGaia19}, grouped by the  year of the corresponding GSA alert. For each solution we report the event time $t_0$, impact parameter $u_0$, Einstein timescale $t_{\rm E}$, parallax components $\pi_{\rm E,N}$ and $\pi_{\rm E,E}$, the Gaia-band source and blend flux parameters, and the $\chi^2 / \rm{dof}$.

The sample is dominated by moderately long events with measurable annual-parallax signatures. Figure~\ref{fig:microlensingParams_distrib} shows the distributions of the main microlensing parameters. The impact parameters are typically small, with a median $|u_0|=0.15$, indicating that the final sample is biased towards relatively high-magnification events. The Einstein timescales span more than an order of magnitude and have a median $t_{\rm E}\simeq 83 \, \rm{days}$. Short events are less likely to show measurable annual-parallax distortions and were therefore less likely to be retained in the sample.

The parallax amplitudes are shown in Fig.~\ref{fig:microlensingParams_distrib}, and the corresponding vector components are shown in Fig.~\ref{fig:pllxParams_distrib}. The accepted solutions have a median $\pi_{\rm E}=0.31$ but cover a broad range of amplitudes and directions. The distribution of $(\pi_{\rm E,N},\pi_{\rm E,E})$ is not expected to be isotropic as Gaia microlensing events are distributed across the sky. The distribution of microlens parallax amplitudes as a function of Einstein time is shown in Fig.~\ref{fig:tepllxParams_distrib}. The largest parallax amplitudes ($\pi_{\rm E} \gtrsim 0.6$) occur for $t_{\rm E} \lesssim 130$\,d, whereas all solutions with $t_{\rm E} \gtrsim 200$\,d have $\pi_{\rm E} \lesssim 0.5$. For a given lens mass, the combination of a large $\pi_{\rm E}$ and a long $t_{\rm E}$ requires a nearby lens with an unusually low relative proper motion, and such events are rare. Conversely, all solutions with $t_{\rm E} < 50$\,d have $\pi_{\rm E} > 0.15$, reflecting our requirement of a significant parallax detection (Sect.~3.4): short events show a measurable annual-parallax distortion only when $\pi_{\rm E}$ is large. Since $M_{\rm L} = \theta_{\rm E}/(\kappa\pi_{\rm E}) = \mu_{\rm rel}t_{\rm E}/(\kappa\pi_{\rm E})$, events in the lower-right part of Fig.~\ref{fig:tepllxParams_distrib} (long $t_{\rm E}$, small $\pi_{\rm E}$) correspond to the most massive lenses for a given relative proper motion and are the natural candidates for massive or dark lenses.

Gaia observes from a Lissajous orbit around the Sun--Earth L2 point, at $\approx 0.01$\,au from the Earth. Because L2 lies on the Sun--Earth line, Gaia's heliocentric trajectory differs from that of the Earth only by a radial scaling of $\approx 1.01$ and by the Lissajous motion ($\lesssim 0.005$\,au). For the annual-parallax signal this corresponds to a change of $\approx 1\%$ in the effective $\pi_{\rm E}$, well below the statistical uncertainties from the best model solutions. For an observer displaced by $D_\perp$ in the plane of the sky, the lens--source relative position shifts by $\pi_{\rm rel}D_\perp/{\rm au}$, i.e. by $\Delta u \simeq \pi_{\rm E} D_\perp/{\rm au}$ in units of $\theta_{\rm E}$ \citep{1966Refsdal, 1994Gould}, where $D_\perp$ is the Earth--Gaia separation projected onto the sky. As $\Delta u$ is a vector offset, it is an upper limit on the change in the scalar separation $u$. The solutions closest to $\Delta u \approx |u_0|$ are high-magnification events; however, for the events in our sample, the nearest Gaia data epoch lies multiple days from $t_0$, where $u(t) \gg \Delta u$. The space parallax signal from Gaia is therefore not detected in the microlensing events in our sample. We therefore treated all datasets in the geocentric frame. A measurement of the Earth--L2 parallax requires dense Gaia and ground-based coverage close to the peak, as for the short-timescale planetary-mass event KMT-2024-BLG-0792/OGLE-2024-BLG-0516 \citep{Dong2026}, and is best investigated on an event-by-event basis.

\section{Gaia-only versus all-data constraints}
\label{sec:gaiaonlyvsfollowup}
Gaia provides homogeneous all-sky discovery photometry, but its sampling is determined by the scanning law and is substantially sparser than that of dedicated ground-based microlensing surveys.
The nominal Gaia scanning law provides approximately 70 focal-plane transits per source over five years, corresponding to an average of about one transit every 26 days, although the observations are distributed non-uniformly and may occur in closely spaced groups \citep{GaiaCollaboration2016}.

This can limit the recovery of higher-order effects, in particular annual parallax. To quantify the role of follow-up photometry, we compared the PSPL and parallax solutions obtained from the full photometric datasets with corresponding fits that used only the Gaia $G$-band alert light curves. Both sets of fits used the same geocentric formalism. As the Earth--L2 effects are at the per-cent level in $\pi_{\rm E}$ (Sect.~4), the differences discussed below reflect the Gaia sampling rather than the observer location.

For each of the 71 accepted full-data solutions associated with 56 unique events, we performed a Gaia-only refit using the same model family, parameter bounds, and blending treatment as in Sect.~\ref{sec:mlModel}. The MCMC chains were initialised from the corresponding best-fitting full-data solution. This setup tests whether the Gaia data can sustain a known solution once initialised near it, rather than whether a blind Gaia-only search would independently recover the same solution. In total, 51 of the 71 Gaia-only fits converge to physically acceptable solutions, corresponding to 41 of the 56 unique events. The remaining fits either fail to converge to a stable solution or yield posteriors that are not sufficiently constrained by the Gaia data alone. Among the 51 acceptable Gaia-only solutions, eight have parallax amplitudes consistent with $\pi_{\rm E}=0$ at the $1\sigma$ level. A non-zero Gaia-only parallax solution does not necessarily imply recovery of the full-data event geometry. Only 21 of the 51 acceptable Gaia-only solutions recover all three microlensing parameters within their $1\sigma$ uncertainty intervals. The individual parameters for $u_0$, $t_{\rm E}$, and $\pi_{\rm E}$ are recovered in only 31, 29, and 37 cases, respectively. The relatively high recovery fraction for $\pi_{\rm E}$ should be interpreted with caution, because in many cases it is driven by broad uncertainty intervals rather than by a precise parallax measurement.

Figure~\ref{fig:gaiaonly_compare} illustrates this behaviour. The impact parameter $u_0$ is the most robustly recovered of the three quantities, although several outliers remain. The Einstein timescale $t_{\rm E}$ shows larger scatter, with some Gaia-only solutions drifting towards substantially different values from those preferred by the full dataset. The strongest degradation is seen for the parallax  amplitude $|\pi_{\rm E}|$, where Gaia-only solutions are frequently broad, scattered, or displaced from the one-to-one relation. Thus, even when a Gaia-only PSPL fit with parallax formally converges, it often does not recover the same higher-order geometry as the combined Gaia and follow-up solution.

This shows that the physical interpretation of GSA microlensing events strongly depends on complementary photometric coverage. Gaia provides the all-sky discovery channel and, in DR4, the astrometric time series against which these events can be tested. However, dense ground-based photometry is usually required to constrain the photometric model and to obtain the parallax constraints needed for lens mass predictions.

\section{Lens mass inference from photometric solutions}
\label{sec:lensMassinference}
Although we extracted the microlensing parallax vector $\pi_{\rm E}$ from the photometric models, a direct mass measurement requires the angular Einstein radius $\theta_{\rm E}$. Gaia DR4 astrometric microlensing models will provide constraints on $\theta_{\rm E}$ for the most favourable events, such that the mass can be determined. Before such astrometric information is available, the lens mass and distance must be inferred statistically by combining the photometric microlensing parameters with Galactic-population priors and source constraints.

We estimated lens properties using two independent approaches. First, we used \texttt{pyLIMASS} \citep{Bachelet2024}, which combines microlensing observables with stellar-population and isochrone information to infer posterior distributions for the physical lens parameters. Second, we used the approach of \citet{Shin2019} and \citet{Bachelet2022} based on the Besan\c{c}on Galaxy Model (BGM) \citep{RobinBGM,CzekajBGM}, in which simulated lens--source populations along the event line of sight are weighted by their consistency with the measured microlensing parameters.

For each accepted photometric solution, both methods were run in two configurations. In the first configuration, labelled \emph{pyL} or \emph{BGM}, we used only the photometric microlensing constraints and Gaia-band flux information. In the second configuration, labelled \emph{pyL+$D_{\rm S}$} or \emph{BGM+$D_{\rm S}$}, we additionally imposed a source-distance prior. This prior was taken from the photogeometric distance estimate of the corresponding Gaia source in the \citet{GaiaCBJ} catalogue.

The event Gaia19ect should be interpreted with caution. Its Gaia alert light curve shows two apparent baseline levels, near $G\simeq19.5$ mag and $G\simeq20.0$ mag, suggesting possible source variability, contamination, or source confusion in the Gaia photometry. We retained Gaia19ect in the sample because Gaia DR4 will provide an updated and longer Gaia time series, which may clarify the origin of the Gaia baseline offset.

\begin{figure}
    \centering
    \includegraphics[width=\columnwidth]{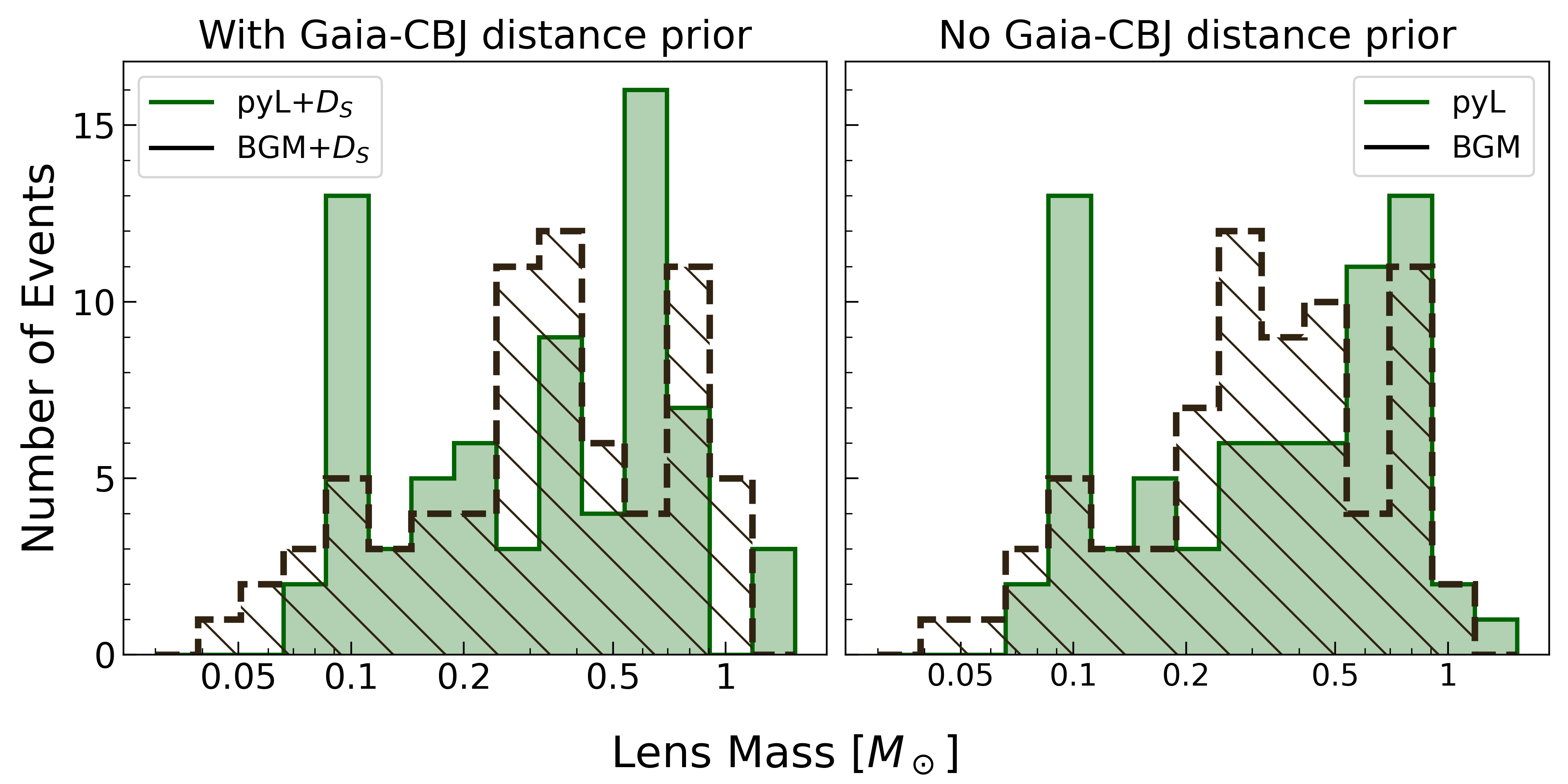}
    \caption{Lens mass distributions inferred with \texttt{pyLIMASS} and the BGM-based Galactic-prior approach. Left panels: Results obtained with the Gaia source-distance prior. Right panels: Corresponding results without this prior.}
    \label{fig:massDistributionPlot}
\end{figure}

\begin{figure}
    \centering
    \includegraphics[width=0.9\columnwidth]{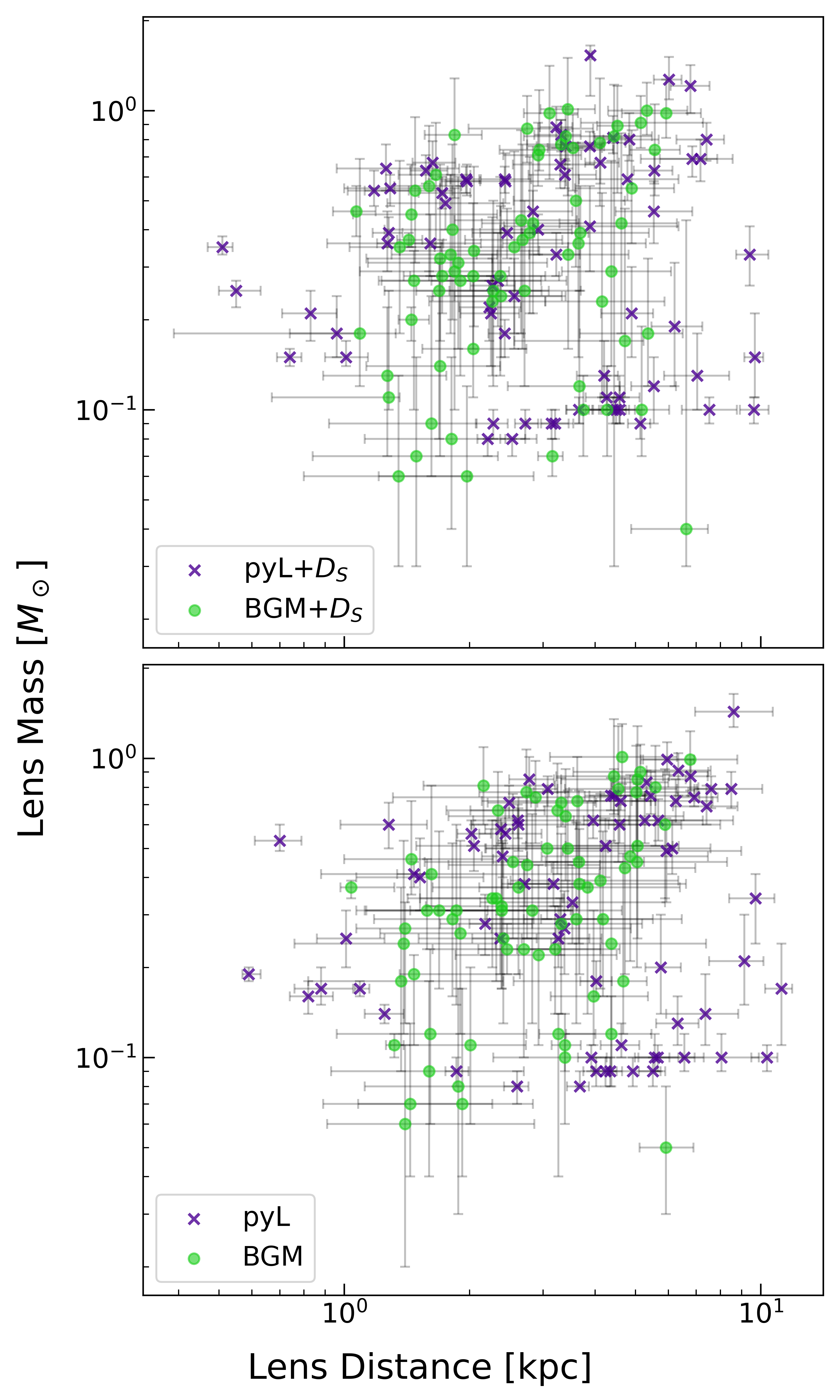}
    \caption{Lens mass and distance distributions inferred with \texttt{pyLIMASS} and the BGM-based Galactic-prior approach. Top panels: Results obtained with the Gaia source-distance prior. Bottom panels: Corresponding results without this prior.}
    \label{fig:massDistanceDistributionPlot}
\end{figure}

\begin{figure}
    \centering
    \includegraphics[width=\columnwidth]{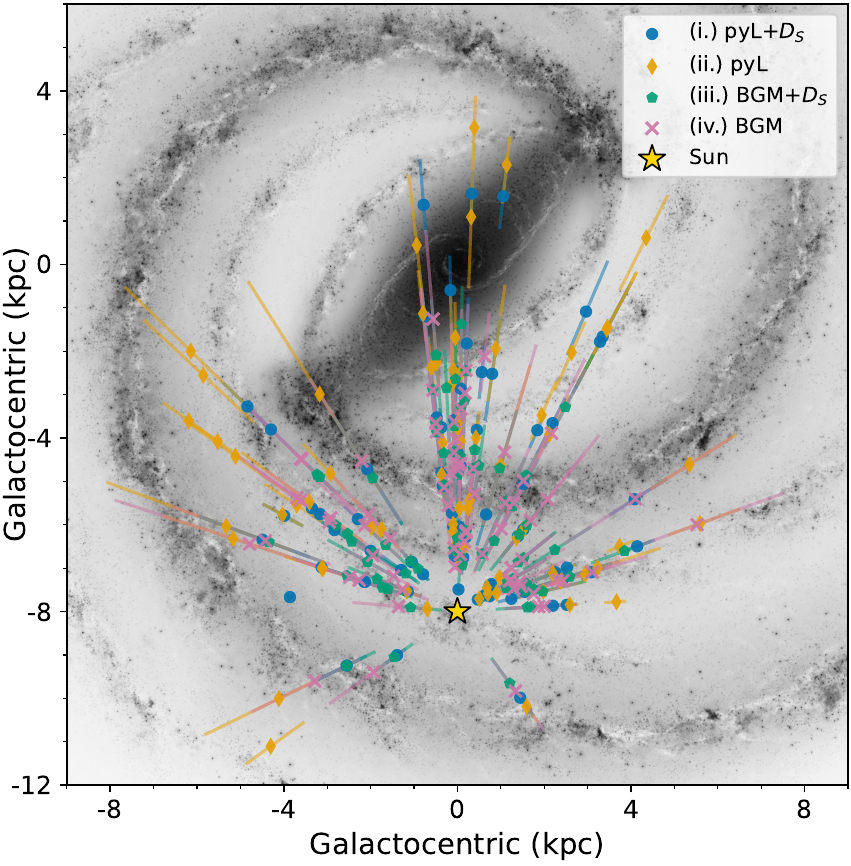}
    \caption{Galactic projection of lens distances inferred with \texttt{pyLIMASS} and the BGM-based Galactic-prior approach. The blue circles and orange diamonds show the lens distance for the \texttt{pyLIMASS} solution where the Gaia source-distance prior was and was not used, respectively. The green pentagon and pink 'x' display the lens distance for the \texttt{BGM}-based Galactic-prior solution with and without the Gaia source-distance prior, respectively. The figure was made using \texttt{MW-Plot} and \emph{Gaia} EDR3 data \citep{GaiaEDR3}.}
    \label{fig:distDistributionPlotONSKY}
\end{figure}

The inferred lens masses and distances are reported in Tables~\ref{tab:lensMassDistGaia17}, ~\ref{tab:lensMassDistGaia18}, and ~\ref{tab:lensMassDistGaia19}, grouped by Gaia alert year. Figure~\ref{fig:massDistributionPlot} shows the resulting lens mass distributions, and Fig.~\ref{fig:massDistanceDistributionPlot} shows the derived lens mass–distance distributions. Figure~\ref{fig:distDistributionPlotONSKY} shows the projected lens positions in the Galactic plane. The spread between the four configurations illustrates the dominant role of population and source-distance priors when $\theta_{\rm E}$ is not yet measured.

None of the inferred lens masses significantly exceeds $\sim 1.5\,M_\odot$. This is largely a consequence of the inference method: without $\theta_{\rm E}$, the mass posterior is dominated by the Galactic and stellar-population priors, in which stellar remnants, and in particular neutron stars and black holes, carry little or no weight, while the lens-flux constraints further disfavour massive luminous lenses. The sample selection also contributes, as the best-studied long-timescale dark-lens candidates among the Gaia alerts (e.g. Gaia18cbf and Gaia18ajz) were excluded from the homogeneous sample. Events with long $t_{\rm E}$ and small $\pi_{\rm E}$, such as Gaia17bdk, Gaia18bmt, Gaia18cik, and Gaia18dif, nevertheless, remain viable candidates for more massive or dark lenses. Because $\theta_{\rm E} = \kappa M_{\rm L}\pi_{\rm E}$, a lens more massive than predicted would produce a proportionally larger astrometric signal, so our DR4 detectability estimates (Sect.~7) are conservative for such lenses, and a Gaia DR4 measurement of $\theta_{\rm E}$ can directly test whether the prior-based masses are underestimated.

\section{Predicted astrometric microlensing signals for Gaia DR4}
\label{sec:predictAstrometry}

The lens mass posteriors from Sect.~\ref{sec:lensMassinference} allow us to predict the astrometric microlensing signal expected in Gaia DR4. We simulated the astrometric signals using the \texttt{astromet}\footnote{\url{https://github.com/zpenoyre/astromet.py}} package \citep{astromet}. For each event, the photometric microlensing parameters were combined with the Gaia DR3 five-parameter astrometric solution, the angular Einstein radius, $\theta_{\rm E}$, inferred from the lens parameters, and the event-specific Gaia scanning law over the DR4 time interval. We used the \texttt{Gaia Observation Forecast Tool}\footnote{https://gaia.esac.esa.int/gost/} to determine Gaia focal-plane crossing times. We then generated Gaia-like along-scan (AL) astrometric measurements for the predicted lensed trajectory. The same simulated data were fitted with a standard five-parameter astrometric model, including position, parallax, and proper motion only. To test whether the Gaia DR4 astrometric time series expected for a known photometric microlensing event should be better described by a standard single-source astrometric solution or by an astrometric microlensing model, we additionally fitted a model extended to include the astrometric microlensing signal.

Uncertainties in the photometric and lens-prior inputs were propagated by Monte Carlo sampling. For each draw, we generated a new astrometric microlensing realisation, simulated the corresponding Gaia AL observations, and refitted both the standard five-parameter and microlensing astrometric models.

A major source of confusion in astrometric microlensing is light from the lens or from unrelated blends within the Gaia window. For a luminous lens with blend-to-source flux ratio $g := f_\mathrm{b} / f_\mathrm{s}$, the centroid shift relative to the unlensed source position becomes $\delta\theta = \theta_{\rm E}\,u\,[A/(u^2+2) - g]/(A+g)$. This reduces to the dark-lens expression $\theta_{\rm E}u/(u^2+2)$ for $g=0$ \citep[e.g.][]{DominikiSahu2000, Jeong1999}. Lens light therefore reduces the amplitude of the shift, alters the shape of the astrometric trajectory, and introduces a term following the lens--source relative motion that can be partly absorbed by the five-parameter solution. Joint fits to the Gaia DR4 epoch photometry and astrometry constrain the lens flux directly, allowing dark and luminous lenses to be distinguished. Because light from an unrelated blend does not follow the lens--source relative motion, its astrometric signature differs from that of a luminous lens. Joint fits can therefore help to establish the nature of the lens and to identify cases in which an additional, unrelated blend must be included in the model.

We quantified the detectability of an astrometric signal in Gaia DR4 using the Bayesian information criterion improvement, $\Delta \mathrm{BIC} = \mathrm{BIC}_{5p} - \mathrm{BIC}_{5p+\mathrm{ML}}$. Positive values indicate that the BIC favours the astrometric-microlensing model over the standard five-parameter (5p) model.

Out of the 284 mass-distance branches derived for the 56 events, we classified 47 branches as high-confidence detections with $\Delta\mathrm{BIC}>10$ and $S/N_{\theta_{\rm E}}>3$. These 47 branches correspond to the following 17 events: Gaia18ajt, Gaia18akb, Gaia18bmt, Gaia18ceb, Gaia18ces, Gaia18cgp, Gaia18cjk, Gaia18cov, Gaia18dep, Gaia18dif, Gaia19aqw, Gaia19bcm, Gaia19bii, Gaia19chi, Gaia19dae, Gaia19dtd, and Gaia19eca. An additional 18 branches have $\Delta\mathrm{BIC}>5$ and $S/N_{\theta_{\rm E}}>3$ and a total of 72 branches have $\Delta\mathrm{BIC}>3$ and $S/N_{\theta_{\rm E}}>3$, which includes the events Gaia18cib, Gaia18cnm, Gaia19eim, and Gaia19eku. For these 72 branches, 61\% (44 branches) recover the injected $\theta_{\rm E}$ value within $1\sigma$.  

Combining a Gaia DR4 measurement of $\theta_{\rm E}$ with the photometric $\pi_{\rm E}$ yields the lens mass directly, $M_{\rm L} = \theta_{\rm E}/(\kappa\pi_{\rm E})$. For the 17 events with high-confidence branches, the photometric $\pi_{\rm E}$ is measured with fractional uncertainties of $< 30 \%$, with the exception of Gaia18bmt. To quantify the expected precision, we propagated, for each simulated realisation, the recovered $\theta_{\rm E}$ and its uncertainty together with the photometric $\pi_{\rm E}$ posterior of the corresponding branch. The median fractional lens mass uncertainty is $24\%$. Since these branches satisfy $S/N_{\theta_{\rm E}} > 3$, the uncertainty on $\theta_{\rm E}$ dominates the mass error budget for most events. The expected precision improves for brighter sources with larger $S/N_{\theta_{\rm E}}$ (Fig.~10). The lens distance, $D_{\rm L} = {\rm au}/(\pi_{\rm rel} + \pi_{\rm S})$ with $\pi_{\rm rel} = \theta_{\rm E}\pi_{\rm E}$, additionally depends on the source parallax $\pi_{\rm S}$, which will be determined directly by joint fits with Gaia DR4 astrometry. Additionally, the astrometric trajectory constrains the direction of $\boldsymbol{\mu}_{\rm rel}$, which must be parallel to $\boldsymbol{\pi}_{\rm E}$. This provides a consistency test of the photometric solution and a means to break discrete degeneracies, such as those between solutions with opposite signs of $u_0$.

\begin{figure}
    \centering
    \includegraphics[width=\columnwidth]{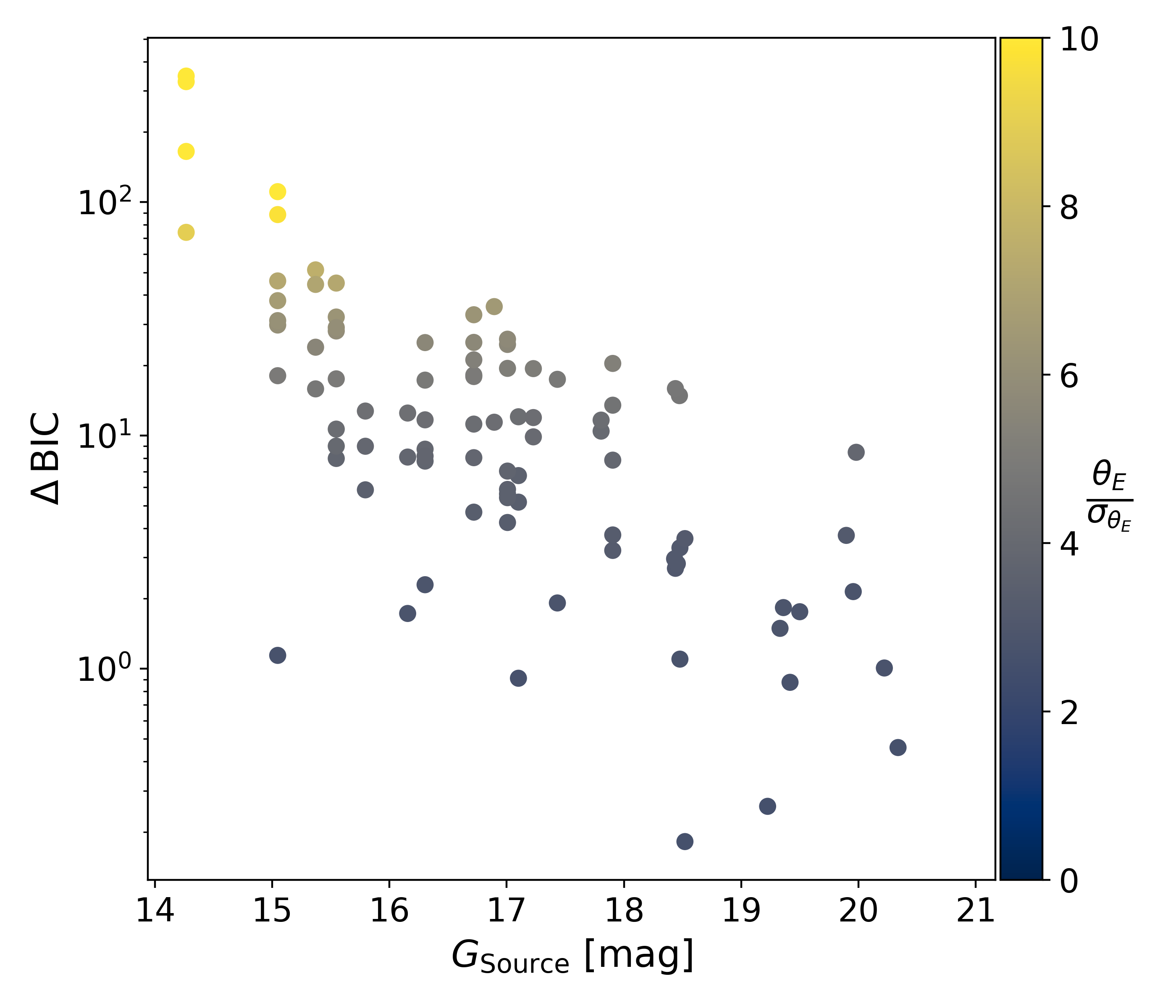}
    \caption{Strength of astrometric microlensing signal as a function of the Gaia DR3 G-band magnitude. The colour bar shows the signal-to-noise ratio of the angular Einstein radius.}
    \label{fig:astrometricPlot}
\end{figure}

We show in Fig.~\ref{fig:astrometricPlot} that the strength of an astrometric signal heavily depends on the brightness of the source. An effective Gaia magnitude limit for astrometric microlensing is not fixed, as it also depends on the angular Einstein radius, event duration, and scanning-law coverage. Astrometric microlensing signals can still be recovered for faint sources when the event has a large angular Einstein radius, $\theta_{\rm E}$ \citep{Jankovic2025}.

\begin{figure}
    \centering
    \includegraphics[width=\columnwidth]{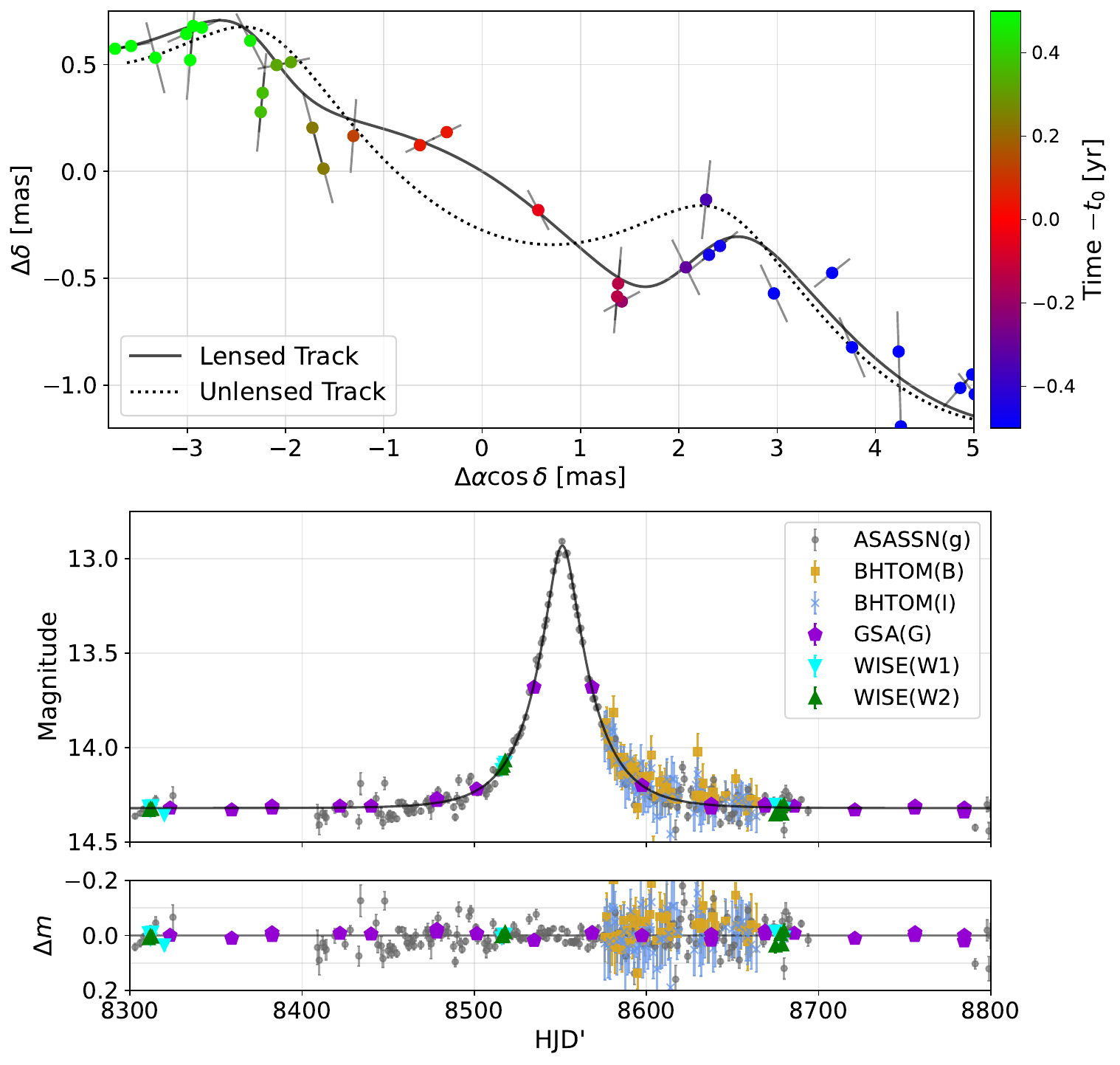}
    \caption{Observed and modelled light curve (bottom) and simulated astrometric track (top) for the Gaia19bcm microlensing event. The astrometric track was simulated for solutions assuming lens parameters from the pyL+$D_S$ solution. The solid line marks the astrometric track due to the microlensing event. The coloured dots mark the simulated Gaia DR4 AL light centroids from the Gaia scanning laws. The dashed line marks the best 5p-fit for the data points. The light curve shows the combined Gaia and follow-up data as well as the combined PSPL and parallax model with parameters from Table~\ref{tab:modelParamsGaia19}.}
    \label{fig:Gaia19bcmPlot}
\end{figure}

\begin{figure}
    \centering
    \includegraphics[width=\columnwidth]{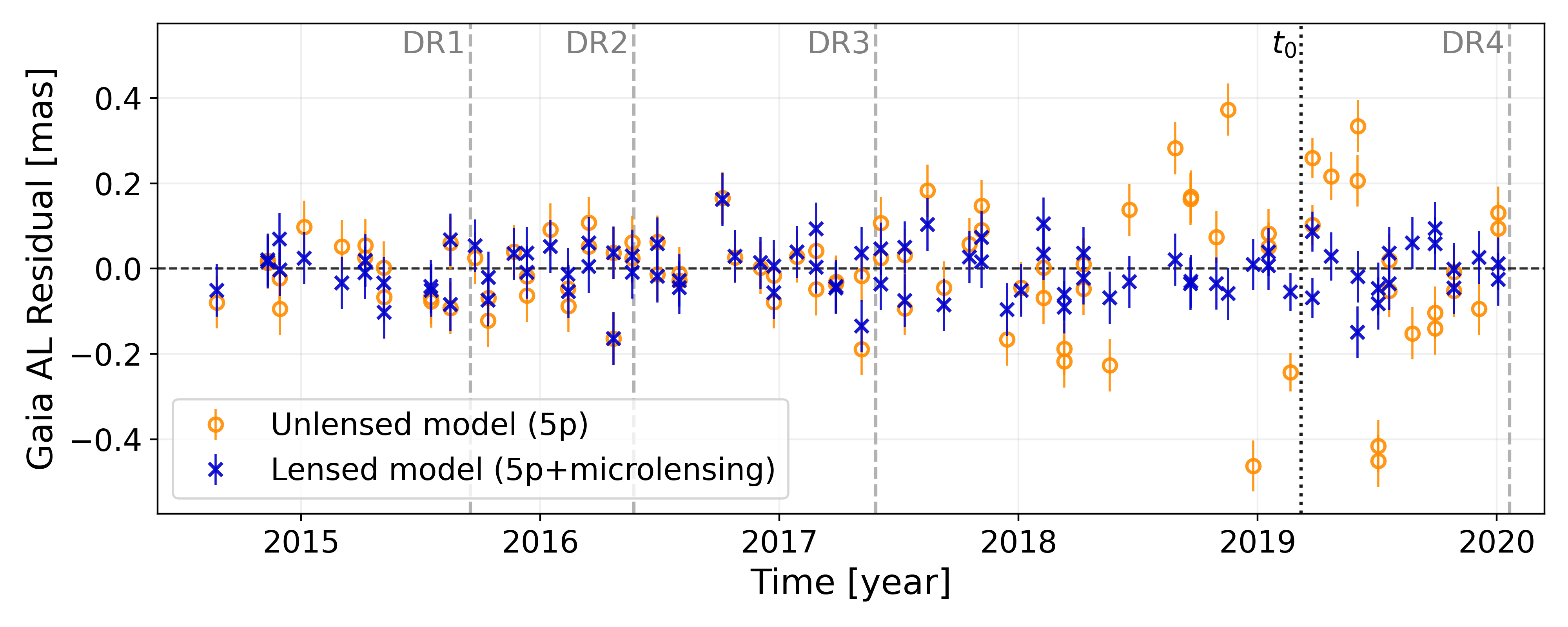}
    \caption{Simulated Gaia AL astrometric residuals for Gaia19bcm, assuming lens parameters from the pyL + $D_S$ solution. The open orange circles show the residuals obtained after fitting a standard five-parameter (5p) astrometric model, while the blue crosses show the residuals obtained with a five-parameter model including astrometric microlensing. The error bars indicate the adopted single-epoch AL uncertainties, and the horizontal dashed line marks the zero residual. The vertical dashed grey lines mark the temporal limits associated with Gaia DR1, DR2, DR3, and DR4, while the dotted black line indicates $t_0$ from the microlensing model.}
    \label{fig:Gaia19bcmALresiduals}
\end{figure}

In Fig.~\ref{fig:Gaia19bcmPlot} we show the light curve and simulated astrometric tracks for Gaia19bcm using the \emph{pyL}+$D_{\rm{S}}$ solution with the source-distance prior. Because Gaia provides 1D AL astrometric measurements, we show the simulated astrometric residuals from the 5p fits with and without microlensing in Fig.~\ref{fig:Gaia19bcmALresiduals}.

Overall, these results indicate that a non-negligible subset of GSA photometric microlensing events should produce detectable astrometric perturbations in Gaia DR4-like data. The most significant cases provide both a statistically preferred astrometric microlensing solution and a meaningful recovery of $\theta_{\rm E}$. This makes them good candidates for combining Gaia astrometry with existing photometric microlensing constraints to infer physical lens properties.

\section{Conclusions}
\label{sec:conclusions}

We constructed a pre-DR4 benchmark sample of GSA microlensing events for which forthcoming Gaia astrometry will be available. The epoch astrometry data in Gaia DR4 can test the photometric microlensing models as well as predicted lens masses and distances.
The GSA archive, constrained to alerts issued before the Gaia DR4 time cut, provided our parent sample of 11\,048 light curves. We first identified candidate microlensing events through Gaia-only PSPL screening, cross-matching with external microlensing and contaminant catalogues, and visual inspection of the Gaia and follow-up light curves. We then performed homogeneous blended PSPL and parallax modelling and retained events for which the photometry supports a non-zero parallax signal. The final sample contains 56 events with 71 accepted PSPL and parallax solutions.

The comparison in Fig.~\ref{fig:gaiaonly_compare} shows that follow-up photometry is vital for the recovery of the microlensing parameters. Gaia provides an all-sky discovery channel and identifies events that are highly valuable for later astrometric verification, but dense external photometry is often required to constrain the photometric solution and to recover physical parameters. This remains an important part of microlensing detections for the LSST \citep{IvezicLSST} alert stream. Despite a median inter-night cadence of approximately three days in the Wide--Fast--Deep survey, coordinated follow-up and cross-survey photometry will still be required for the robust classification and modelling of microlensing events \citep{AbramsLSST}. Spectroscopic follow-up, even after the microlensing event has ended, provides complementary constraints by improving the characterisation of the source star. These include constraints on its effective temperature, surface gravity, metallicity, and distance, which can in turn strengthen the inference of the physical lens properties.

Using \texttt{pyLIMASS} and a BGM-based Galactic-prior framework, we derived lens mass and distance predictions for each accepted photometric solution. The differences between the inference configurations show that, without a direct measurement of $\theta_{\rm E}$, the inferred lens properties remain strongly prior-dependent. Our Gaia DR4-like simulations predict that 47 event--solution branches should be high-confidence astrometric detections, with a further 25 potentially detectable cases. This motivates the use of Gaia DR4 astrometry, and other complementary constraints, to turn these photometric predictions into physical lens measurements. 

\begin{acknowledgements}
BHTOM.space is based on the open-source TOM Toolkit by LCO and has been developed with funding from the OPTICON-RadioNet Pilot (ORP) of the European Union's Horizon 2020 research and innovation programme under grant agreement No 101004719 (2021-2025). This project has received funding from the European Union's Horizon Europe Research and Innovation programme ACME under grant agreement No 101131928 (2024-2028). {\L}.W. acknowledges support from the Polish National Science Centre DAINA grant No 2024/52/L/ST9/00210. E.S., M.M., E.P., J.Z., V.C., and U.J. acknowledge funding from the Research Council of Lithuania (LMTLT, grant No. S-LL-24-1).
This work makes use of observations from the Las Cumbres Observatory global telescope network.
This work has made use of data from the European Space Agency (ESA) mission \textit{Gaia} (\url{https://www.cosmos.esa.int/gaia}), processed by the \textit{Gaia} Data Processing and Analysis Consortium (DPAC; \url{https://www.cosmos.esa.int/web/gaia/dpac/consortium}). Funding for DPAC has been provided by national institutions, in particular the institutions participating in the \textit{Gaia} Multilateral Agreement.
This research has made use of publicly available data (https://kmtnet.kasi.re.kr/ulens/) from the KMTNet system operated by the Korea Astronomy and Space Science Institute (KASI) at three host sites of CTIO in Chile, SAAO in South Africa, and SSO in Australia. Data transfer from the host site to KASI was supported by the Korea Research Environment Open NETwork (KREONET).
This paper makes use of data obtained by the MOA collaboration with the 1.8-metre MOA-II telescope at the University of Canterbury Mount John Observatory, Lake Tekapo, New Zealand. The MOA project is supported by JSPS KAKENHI Grant Number JP16H06287, JP22H00153, JP23KK0060 and JP25H00668. These data are made available using services at the NASA Exoplanet Archive, which is operated by the California Institute of Technology, under contract with the National Aeronautics and Space Administration under the Exoplanet Exploration Program.
This work makes use of observations obtained with the Samuel Oschin Telescope 48-inch and the 60-inch Telescope at the Palomar Observatory as part of the Zwicky Transient Facility project. ZTF is supported by the National Science Foundation under Award 2407588 and a partnership including Caltech, USA; Caltech/IPAC, USA; University of Maryland, USA; University of California, Berkeley, USA; University of Wisconsin at Milwaukee, USA; Cornell University, USA; Drexel University, USA; University of North Carolina at Chapel Hill, USA; Institute of Science and Technology, Austria; National Central University, Taiwan; and OKC, University of Stockholm, Sweden. Operations are conducted by Caltech’s Optical Observatory (COO), Caltech/IPAC, and the University of Washington at Seattle, USA.
JMC was (partially) supported by the Spanish MICIN/AEI/10.13039/501100011033 and by "ERDF A way of making Europe" by the "European Union" through grant PID2021-122842OB-C21 and PID2024-157964OB-C21, and the Institute of Cosmos Sciences University of Barcelona (ICCUB, Unidad de Excelencia 'Mar\'{\i}a de Maeztu') through grant CEX2024-001451-M and the project 2021-SGR-00679 GRC de l'Agència de Gestió d'Ajuts Universitaris i de Recerca (Generalitat de Catalunya). The Joan Oró Telescope (TJO) of the Montsec Observatory (OdM) is owned by the Catalan Government and operated by the Institute for Space Studies of Catalonia (IEEC).
H.H.E. thanks the Türkiye National Observatories for partial support in using the TUG-T100 telescope under project number 15CT100-915.
M.W. thanks M. J. Mr\'{o}z for his collaboration and feedback.
\end{acknowledgements}

\bibliographystyle{aa}
\bibliography{bibs}

\clearpage
\onecolumn

\begin{appendix}
\section{Model parameters}

\begin{table*}[!htbp]
\caption{Model parameters for GSA alerts from 2017.}
\label{tab:modelParamsGaia17}
\centering
\footnotesize
\setlength{\tabcolsep}{1pt}
\renewcommand{\arraystretch}{1.2}
\begin{tabular*}{\textwidth}{@{\extracolsep{\fill}}lcccccccccc}
\toprule
Event &
RA (\textdegree) &
Dec (\textdegree) &
$t_0$ (HJD') &
$u_0$ &
$t_E$ (days) &
$\pi_{E,N}$ &
$\pi_{E,E}$ &
$f_{\mathrm{b}}/f_{\mathrm{s}}$ &
$\chi^2$ \, / \, dof &
Mod. \\
\midrule
Gaia17aqu & 116.0932 & -28.4434 & $7880.592_{-0.155}^{+0.174}$ & $-0.080_{-0.007}^{+0.008}$ & $89.47_{-5.54}^{+7.69}$ & $0.196_{-0.058}^{+0.040}$ & $0.050_{-0.029}^{+0.020}$ & $0.15_{-0.10}^{+0.14}$ & 144.3 / 264 & A \\
Gaia17bbs & 272.7051 & -22.6155 & $7861.190_{-0.592}^{+0.117}$ & $0.013_{-0.022}^{+0.001}$ & $193.83_{-20.72}^{+7.07}$ & $-0.151_{-0.010}^{+0.288}$ & $-0.082_{-0.010}^{+0.022}$ & $0.06_{-0.22}^{+0.27}$ & 42.0 / 92 & A \\
Gaia17bdk & 265.6345 & -34.1037 & $7878.511_{-0.003}^{+0.003}$ & $0.157_{-0.001}^{+0.001}$ & $56.38_{-0.11}^{+0.12}$ & $-0.038_{-0.021}^{+0.021}$ & $-0.051_{-0.006}^{+0.007}$ & $0.00_{-0.01}^{+0.01}$ & 447.1 / 626 & A \\
Gaia17bej & 274.2544 & -31.3835 & $7875.220_{-0.056}^{+0.080}$ & $0.029_{-0.008}^{+0.009}$ & $118.76_{-10.73}^{+11.65}$ & $-0.039_{-0.124}^{+0.120}$ & $0.179_{-0.034}^{+0.037}$ & $0.11_{-0.15}^{+0.16}$ & 157.4 / 199 & A \\
Gaia17cah & 268.9144 & -31.3086 & $7966.552_{-0.002}^{+0.002}$ & $0.092_{-0.001}^{+0.001}$ & $57.35_{-0.12}^{+0.12}$ & $0.346_{-0.022}^{+0.024}$ & $0.011_{-0.007}^{+0.005}$ & $-0.05_{-0.01}^{+0.01}$ & 2076.9 / 3105 & A \\
\bottomrule
\end{tabular*}
\tablefoot{Microlensing model parameters are reported as $x_{-\sigma_-}^{+\sigma_+}$, with uncertainties corresponding to the 16th and 84th percentiles.}
\end{table*}

\begin{table*}[!htbp]
\caption{Model parameters for GSA alerts from 2018.}
\label{tab:modelParamsGaia18}
\centering
\footnotesize
\setlength{\tabcolsep}{1pt}
\renewcommand{\arraystretch}{1.2}
\begin{tabular*}{\textwidth}{@{\extracolsep{\fill}}lcccccccccc}
\toprule
Event &
RA (\textdegree) &
Dec (\textdegree) &
$t_0$ (HJD') &
$u_0$ &
$t_E$ (days) &
$\pi_{E,N}$ &
$\pi_{E,E}$ &
$f_{\mathrm{b}}/f_{\mathrm{s}}$ &
$\chi^2$ \, / \, dof &
Mod. \\
\midrule
Gaia18ajt & 267.1333 & -35.0880 & $8140.297_{-0.028}^{+0.027}$ & $0.172_{-0.001}^{+0.002}$ & $42.19_{-0.27}^{+0.21}$ & $-0.335_{-0.019}^{+0.016}$ & $0.235_{-0.010}^{+0.013}$ & $-0.14_{-0.03}^{+0.03}$ & 3565.9 / 3539 & A \\
Gaia18akb & 270.6288 & -27.8430 & $8178.376_{-0.050}^{+0.025}$ & $0.516_{-0.021}^{+0.035}$ & $48.97_{-2.32}^{+1.74}$ & $0.779_{-0.017}^{+0.023}$ & $0.175_{-0.019}^{+0.023}$ & $1.37_{-0.57}^{+0.44}$ & 28280.2 / 4057 & A \\
Gaia18bmt & 214.0148 & -56.9134 & $8284.806_{-0.030}^{+0.029}$ & $0.068_{-0.001}^{+0.001}$ & $60.37_{-0.83}^{+1.04}$ & $0.003_{-0.036}^{+0.033}$ & $0.032_{-0.019}^{+0.022}$ & $0.15_{-0.02}^{+0.02}$ & 1462.7 / 874 & A \\
 &  &  & $8284.803_{-0.027}^{+0.033}$ & $-0.068_{-0.002}^{+0.001}$ & $60.41_{-0.95}^{+1.17}$ & $0.007_{-0.043}^{+0.032}$ & $0.033_{-0.019}^{+0.023}$ & $0.15_{-0.02}^{+0.02}$ & 1462.7 / 874 & B \\
Gaia18ceb & 264.8469 & -35.2672 & $8302.324_{-0.059}^{+0.070}$ & $-0.909_{-0.058}^{+0.032}$ & $49.82_{-1.96}^{+1.33}$ & $0.394_{-0.171}^{+0.138}$ & $0.260_{-0.028}^{+0.026}$ & $-0.16_{-0.28}^{+0.25}$ & 717.5 / 736 & A \\
Gaia18cer & 263.9740 & -27.5328 & $8375.523_{-0.016}^{+0.017}$ & $-0.434_{-0.006}^{+0.005}$ & $43.85_{-0.45}^{+0.43}$ & $0.004_{-0.033}^{+0.034}$ & $0.458_{-0.009}^{+0.010}$ & $-0.12_{-0.02}^{+0.02}$ & 204.0 / 643 & A \\
 &  &  & $8375.536_{-0.015}^{+0.019}$ & $0.435_{-0.005}^{+0.005}$ & $44.29_{-0.41}^{+0.50}$ & $0.049_{-0.026}^{+0.036}$ & $0.454_{-0.009}^{+0.009}$ & $-0.12_{-0.02}^{+0.02}$ & 203.4 / 643 & B \\
Gaia18ces & 269.2658 & -33.9361 & $8450.954_{-0.316}^{+0.415}$ & $0.898_{-0.049}^{+0.115}$ & $32.54_{-2.02}^{+0.96}$ & $-0.175_{-0.022}^{+0.011}$ & $0.536_{-0.015}^{+0.035}$ & $-0.83_{-0.08}^{+0.06}$ & 147774.0 / 1897 & A \\
Gaia18cfc & 268.5731 & -30.5694 & $8320.653_{-0.004}^{+0.004}$ & $0.221_{-0.001}^{+0.001}$ & $44.10_{-0.11}^{+0.13}$ & $-0.079_{-0.061}^{+0.064}$ & $0.267_{-0.019}^{+0.018}$ & $-0.00_{-0.02}^{+0.02}$ & 931.3 / 2882 & A \\
Gaia18cfs & 268.6784 & -22.8204 & $8348.277_{-0.001}^{+0.001}$ & $-0.022_{-0.001}^{+0.001}$ & $40.98_{-0.16}^{+0.18}$ & $-0.021_{-0.070}^{+0.068}$ & $0.168_{-0.008}^{+0.008}$ & $-0.12_{-0.01}^{+0.01}$ & 1355.5 / 731 & A \\
Gaia18cgp & 271.7968 & -24.9823 & $8315.840_{-0.001}^{+0.001}$ & $-0.009_{-0.001}^{+0.001}$ & $45.40_{-0.04}^{+0.04}$ & $0.353_{-0.052}^{+0.042}$ & $-0.063_{-0.005}^{+0.006}$ & $0.16_{-0.01}^{+0.01}$ & 6037.7 / 743 & A \\
Gaia18cib & 279.1148 & -16.5783 & $8327.392_{-0.085}^{+0.152}$ & $-0.004_{-0.021}^{+0.022}$ & $77.18_{-9.82}^{+5.52}$ & $-0.397_{-0.423}^{+0.145}$ & $-0.250_{-0.116}^{+0.102}$ & $0.06_{-0.18}^{+0.10}$ & 56.4 / 85 & A \\
Gaia18cik & 278.3799 & -3.4032 & $8364.093_{-1.194}^{+2.302}$ & $0.212_{-0.026}^{+0.067}$ & $211.34_{-33.01}^{+16.62}$ & $0.103_{-0.017}^{+0.022}$ & $-0.191_{-0.067}^{+0.034}$ & $-0.20_{-0.23}^{+0.13}$ & 393.0 / 118 & A \\
 &  &  & $8362.627_{-0.784}^{+1.043}$ & $-0.084_{-0.028}^{+0.008}$ & $396.85_{-96.29}^{+46.43}$ & $0.078_{-0.022}^{+0.016}$ & $-0.061_{-0.014}^{+0.008}$ & $1.30_{-0.60}^{+0.26}$ & 400.8 / 118 & B \\
Gaia18cjk & 283.7880 & -6.6635 & $8377.218_{-8.601}^{+3.161}$ & $-0.156_{-0.070}^{+0.020}$ & $214.16_{-48.22}^{+23.81}$ & $-0.343_{-0.144}^{+0.087}$ & $0.360_{-0.037}^{+0.090}$ & $2.63_{-1.20}^{+0.59}$ & 1034.2 / 257 & A \\
Gaia18clg & 155.5824 & -64.5504 & $8378.023_{-0.494}^{+0.699}$ & $0.104_{-0.018}^{+0.034}$ & $150.98_{-30.08}^{+25.29}$ & $0.270_{-0.032}^{+0.056}$ & $0.116_{-0.033}^{+0.041}$ & $0.07_{-0.28}^{+0.24}$ & 63.8 / 203 & A \\
 &  &  & $8378.425_{-0.637}^{+0.583}$ & $-0.107_{-0.027}^{+0.022}$ & $144.80_{-23.62}^{+32.76}$ & $0.279_{-0.036}^{+0.050}$ & $0.210_{-0.052}^{+0.048}$ & $0.04_{-0.22}^{+0.29}$ & 63.7 / 203 & B \\
Gaia18cmk & 307.1494 & 22.8305 & $8362.444_{-0.075}^{+0.064}$ & $0.386_{-0.021}^{+0.020}$ & $31.15_{-1.00}^{+1.35}$ & $-0.504_{-0.075}^{+0.147}$ & $-0.596_{-0.059}^{+0.112}$ & $0.21_{-0.08}^{+0.10}$ & 5617.5 / 894 & A \\
Gaia18cnm & 275.8788 & -7.3327 & $8391.009_{-1.233}^{+0.498}$ & $0.611_{-0.185}^{+0.133}$ & $54.94_{-6.61}^{+33.32}$ & $-0.073_{-0.110}^{+0.495}$ & $-0.299_{-0.106}^{+0.014}$ & $-0.29_{-0.20}^{+0.53}$ & 2691.1 / 898 & A \\
Gaia18cov & 108.6160 & -20.9863 & $8422.671_{-2.474}^{+1.218}$ & $-0.314_{-0.160}^{+0.044}$ & $120.24_{-38.71}^{+20.57}$ & $0.032_{-0.124}^{+0.043}$ & $-0.307_{-0.043}^{+0.020}$ & $1.13_{-0.95}^{+0.47}$ & 1683.7 / 616 & A \\
Gaia18cwh & 264.4138 & -35.3457 & $8417.155_{-0.195}^{+0.090}$ & $-0.001_{-0.006}^{+0.009}$ & $84.64_{-4.92}^{+7.23}$ & $0.049_{-0.099}^{+0.097}$ & $-0.082_{-0.033}^{+0.036}$ & $0.93_{-0.38}^{+0.67}$ & 585.6 / 706 & A \\
Gaia18dep & 135.1505 & -56.5469 & $8419.245_{-0.888}^{+0.723}$ & $-0.160_{-0.026}^{+0.022}$ & $101.80_{-11.74}^{+13.08}$ & $-0.027_{-0.037}^{+0.035}$ & $-0.625_{-0.060}^{+0.062}$ & $5.02_{-0.99}^{+1.05}$ & 644.8 / 966 & A \\
Gaia18dif & 197.5516 & -58.8041 & $8525.322_{-0.088}^{+0.066}$ & $-0.176_{-0.009}^{+0.007}$ & $215.09_{-9.75}^{+9.52}$ & $-0.086_{-0.007}^{+0.006}$ & $-0.052_{-0.005}^{+0.004}$ & $0.98_{-0.11}^{+0.11}$ & 1681.6 / 740 & A \\
 &  &  & $8524.862_{-0.144}^{+0.137}$ & $0.266_{-0.013}^{+0.015}$ & $198.06_{-7.40}^{+7.26}$ & $-0.079_{-0.005}^{+0.005}$ & $-0.035_{-0.003}^{+0.003}$ & $0.19_{-0.08}^{+0.08}$ & 1637.0 / 740 & B \\
Gaia18dma & 295.3575 & 13.8726 & $8445.942_{-0.157}^{+0.169}$ & $-0.079_{-0.009}^{+0.007}$ & $107.60_{-9.74}^{+9.36}$ & $0.278_{-0.053}^{+0.055}$ & $-0.757_{-0.043}^{+0.038}$ & $0.52_{-0.17}^{+0.17}$ & 7648.2 / 237 & A \\
 &  &  & $8445.085_{-0.315}^{+0.249}$ & $0.080_{-0.009}^{+0.008}$ & $125.29_{-11.69}^{+15.59}$ & $0.215_{-0.059}^{+0.058}$ & $-0.849_{-0.047}^{+0.045}$ & $0.48_{-0.15}^{+0.19}$ & 7651.7 / 237 & B \\
Gaia18dml & 292.1442 & 23.1339 & $8495.592_{-0.692}^{+1.181}$ & $0.166_{-0.201}^{+0.561}$ & $95.74_{-48.81}^{+31.79}$ & $0.084_{-0.021}^{+0.087}$ & $0.209_{-0.074}^{+0.309}$ & $3.95_{-3.94}^{+2.80}$ & 2655.2 / 1281 & A \\
Gaia18eaz & 219.3758 & -67.2788 & $8497.233_{-0.594}^{+2.745}$ & $-0.075_{-0.003}^{+0.067}$ & $39.32_{-4.45}^{+3.53}$ & $-0.386_{-0.169}^{+0.399}$ & $-0.194_{-0.178}^{+0.043}$ & $-0.17_{-0.25}^{+0.13}$ & 100.3 / 297 & A \\
\bottomrule
\end{tabular*}
\tablefoot{Microlensing model parameters are reported as $x_{-\sigma_-}^{+\sigma_+}$, with uncertainties corresponding to the 16th and 84th percentiles.}
\end{table*}

\begin{table*}[!htbp]
\caption{Model parameters for GSA alerts from 2019.}
\label{tab:modelParamsGaia19}
\centering
\footnotesize
\setlength{\tabcolsep}{1pt}
\renewcommand{\arraystretch}{1.2}
\begin{tabular*}{\textwidth}{@{\extracolsep{\fill}}lcccccccccc}
\toprule
Event &
RA (\textdegree) &
Dec (\textdegree) &
$t_0$ (HJD') &
$u_0$ &
$t_E$ (days) &
$\pi_{E,N}$ &
$\pi_{E,E}$ &
$f_{\mathrm{b}}/f_{\mathrm{s}}$ &
$\chi^2$ \, / \, dof &
Mod. \\
\midrule
Gaia19aqf & 287.9695 & 3.9495 & $8545.968_{-0.470}^{+1.408}$ & $-0.113_{-0.150}^{+0.011}$ & $209.69_{-101.05}^{+18.27}$ & $-0.258_{-0.018}^{+0.036}$ & $0.080_{-0.018}^{+0.017}$ & $3.03_{-2.57}^{+0.47}$ & 795.4 / 656 & A \\
 &  &  & $8545.042_{-1.362}^{+1.357}$ & $0.332_{-0.056}^{+0.361}$ & $110.73_{-44.71}^{+13.25}$ & $-0.319_{-0.035}^{+0.059}$ & $0.277_{-0.039}^{+0.103}$ & $0.07_{-0.71}^{+0.23}$ & 794.2 / 656 & B \\
Gaia19aqw & 182.2558 & -68.0368 & $8516.198_{-0.046}^{+0.047}$ & $-0.060_{-0.002}^{+0.002}$ & $81.46_{-1.31}^{+1.20}$ & $-0.276_{-0.008}^{+0.007}$ & $-0.257_{-0.008}^{+0.006}$ & $-0.04_{-0.02}^{+0.02}$ & 313.2 / 704 & A \\
Gaia19asx & 289.1144 & 26.6535 & $8566.298_{-0.054}^{+0.044}$ & $0.109_{-0.004}^{+0.003}$ & $40.09_{-0.77}^{+0.99}$ & $0.953_{-0.036}^{+0.023}$ & $0.283_{-0.019}^{+0.015}$ & $0.62_{-0.05}^{+0.06}$ & 5221.2 / 1168 & A \\
 &  &  & $8566.295_{-0.057}^{+0.055}$ & $-0.107_{-0.004}^{+0.003}$ & $42.58_{-1.00}^{+0.95}$ & $0.878_{-0.029}^{+0.029}$ & $0.260_{-0.017}^{+0.016}$ & $0.65_{-0.06}^{+0.06}$ & 5223.9 / 1168 & B \\
Gaia19awc & 273.9006 & -2.2570 & $8612.638_{-0.118}^{+0.136}$ & $0.314_{-0.021}^{+0.021}$ & $64.62_{-3.89}^{+4.56}$ & $-0.038_{-0.119}^{+0.116}$ & $-0.075_{-0.027}^{+0.025}$ & $0.00_{-0.09}^{+0.10}$ & 3949.5 / 1107 & A \\
Gaia19ayo & 265.2167 & -36.0396 & $8534.788_{-0.007}^{+0.015}$ & $0.062_{-0.002}^{+0.002}$ & $57.66_{-1.95}^{+1.36}$ & $-0.385_{-0.012}^{+0.539}$ & $0.323_{-0.074}^{+0.008}$ & $0.25_{-0.03}^{+0.04}$ & 169.5 / 617 & A \\
 &  &  & $8534.791_{-0.008}^{+0.012}$ & $-0.063_{-0.001}^{+0.002}$ & $56.40_{-1.08}^{+1.76}$ & $0.155_{-0.550}^{+0.015}$ & $0.284_{-0.029}^{+0.042}$ & $0.23_{-0.02}^{+0.05}$ & 169.5 / 617 & B \\
Gaia19bcm & 156.8948 & -71.7010 & $8551.198_{-0.038}^{+0.030}$ & $-0.271_{-0.011}^{+0.011}$ & $31.93_{-1.04}^{+1.16}$ & $-0.174_{-0.035}^{+0.034}$ & $0.498_{-0.134}^{+0.129}$ & $0.07_{-0.05}^{+0.06}$ & 2300.8 / 1380 & A \\
Gaia19bht & 268.8967 & -29.0048 & $8574.830_{-0.015}^{+0.012}$ & $-0.319_{-0.003}^{+0.003}$ & $54.98_{-0.42}^{+0.51}$ & $-0.042_{-0.043}^{+0.048}$ & $-0.267_{-0.003}^{+0.004}$ & $1.60_{-0.09}^{+0.10}$ & 4128.3 / 709 & A \\
Gaia19bhw & 267.3231 & -30.8515 & $8589.946_{-0.372}^{+0.280}$ & $0.301_{-0.034}^{+0.050}$ & $78.09_{-8.85}^{+7.91}$ & $-0.428_{-0.080}^{+0.085}$ & $0.042_{-0.022}^{+0.031}$ & $1.45_{-1.11}^{+1.63}$ & 127.5 / 381 & A \\
Gaia19bii & 269.5016 & -33.7132 & $8554.110_{-0.002}^{+0.001}$ & $0.062_{-0.001}^{+0.001}$ & $34.16_{-0.04}^{+0.04}$ & $0.298_{-0.019}^{+0.019}$ & $0.004_{-0.003}^{+0.002}$ & $0.03_{-0.01}^{+0.01}$ & 6381.9 / 1234 & A \\
Gaia19bir & 234.6434 & -51.2415 & $8612.348_{-2.458}^{+9.933}$ & $-0.066_{-0.269}^{+0.205}$ & $128.82_{-76.79}^{+14.81}$ & $0.011_{-0.021}^{+0.107}$ & $-0.289_{-0.444}^{+0.040}$ & $0.29_{-1.10}^{+0.15}$ & 8.6 / 41 & A \\
Gaia19bjq & 55.1976 & 57.9558 & $8618.113_{-0.228}^{+0.220}$ & $-0.121_{-0.013}^{+0.014}$ & $132.69_{-3.62}^{+3.40}$ & $-0.214_{-0.005}^{+0.003}$ & $0.217_{-0.004}^{+0.007}$ & $0.06_{-0.05}^{+0.05}$ & 2202.2 / 1449 & A \\
Gaia19brt & 297.7061 & 34.6373 & $8650.944_{-0.006}^{+0.007}$ & $0.015_{-0.001}^{+0.001}$ & $103.66_{-2.24}^{+2.45}$ & $-0.046_{-0.015}^{+0.011}$ & $-0.103_{-0.014}^{+0.019}$ & $6.94_{-0.21}^{+0.23}$ & 7124.0 / 1555 & A \\
Gaia19chi & 232.0353 & -67.0209 & $8637.369_{-1.235}^{+1.496}$ & $0.221_{-0.082}^{+0.116}$ & $133.68_{-37.22}^{+63.71}$ & $0.110_{-0.029}^{+0.040}$ & $0.185_{-0.047}^{+0.020}$ & $1.60_{-1.11}^{+1.97}$ & 362.3 / 240 & A \\
 &  &  & $8637.061_{-1.176}^{+2.727}$ & $-0.330_{-0.154}^{+0.156}$ & $101.00_{-27.55}^{+83.75}$ & $0.302_{-0.074}^{+0.067}$ & $0.131_{-0.077}^{+0.063}$ & $0.54_{-0.66}^{+1.95}$ & 361.8 / 240 & B \\
Gaia19cyv & 318.2633 & 43.3376 & $8728.572_{-0.036}^{+0.037}$ & $0.453_{-0.015}^{+0.018}$ & $48.33_{-1.37}^{+1.23}$ & $-0.566_{-0.021}^{+0.018}$ & $-0.418_{-0.019}^{+0.018}$ & $-0.13_{-0.05}^{+0.04}$ & 6036.4 / 1116 & A \\
 &  &  & $8728.913_{-0.053}^{+0.054}$ & $-0.461_{-0.018}^{+0.017}$ & $56.20_{-1.37}^{+1.37}$ & $-0.474_{-0.016}^{+0.016}$ & $-0.268_{-0.009}^{+0.009}$ & $-0.16_{-0.05}^{+0.05}$ & 5934.7 / 1116 & B \\
Gaia19dae & 309.0341 & 32.7209 & $8720.933_{-0.172}^{+0.306}$ & $0.467_{-0.040}^{+0.069}$ & $119.20_{-14.89}^{+11.42}$ & $0.204_{-0.021}^{+0.011}$ & $0.119_{-0.021}^{+0.014}$ & $0.70_{-0.32}^{+0.24}$ & 14915.3 / 2548 & A \\
 &  &  & $8721.314_{-0.127}^{+0.108}$ & $-0.404_{-0.037}^{+0.028}$ & $107.34_{-6.48}^{+5.82}$ & $0.108_{-0.006}^{+0.004}$ & $0.115_{-0.010}^{+0.007}$ & $1.10_{-0.25}^{+0.23}$ & 14890.7 / 2548 & B \\
Gaia19dak & 302.3652 & 29.9359 & $8898.094_{-1.078}^{+0.892}$ & $0.397_{-0.007}^{+0.009}$ & $216.24_{-3.47}^{+3.17}$ & $-0.199_{-0.005}^{+0.004}$ & $0.436_{-0.005}^{+0.007}$ & $-0.08_{-0.02}^{+0.02}$ & 10803.1 / 1524 & A \\
 &  &  & $8949.083_{-0.813}^{+1.031}$ & $0.011_{-0.002}^{+0.004}$ & $268.92_{-3.71}^{+3.64}$ & $-0.043_{-0.003}^{+0.003}$ & $0.401_{-0.006}^{+0.006}$ & $0.02_{-0.02}^{+0.02}$ & 10822.8 / 1524 & B \\
Gaia19drc & 262.8726 & -30.3847 & $8720.391_{-0.077}^{+0.078}$ & $0.329_{-0.023}^{+0.025}$ & $80.35_{-4.13}^{+4.46}$ & $-0.425_{-0.021}^{+0.027}$ & $0.170_{-0.014}^{+0.015}$ & $1.70_{-0.70}^{+0.95}$ & 976.1 / 630 & A \\
Gaia19dtd & 272.9919 & -29.9231 & $8693.259_{-0.044}^{+0.046}$ & $-0.613_{-0.014}^{+0.011}$ & $63.80_{-0.67}^{+0.56}$ & $-0.150_{-0.038}^{+0.033}$ & $0.010_{-0.010}^{+0.009}$ & $-0.01_{-0.04}^{+0.03}$ & 3487.7 / 2782 & A \\
 &  &  & $8693.284_{-0.041}^{+0.046}$ & $0.653_{-0.017}^{+0.021}$ & $60.27_{-1.10}^{+0.97}$ & $0.094_{-0.018}^{+0.019}$ & $-0.012_{-0.011}^{+0.012}$ & $-0.11_{-0.05}^{+0.04}$ & 3488.7 / 2782 & B \\
Gaia19dtk & 280.4294 & -28.4836 & $8829.584_{-6.444}^{+5.523}$ & $-0.217_{-0.075}^{+0.043}$ & $162.62_{-44.22}^{+49.89}$ & $0.427_{-0.088}^{+0.134}$ & $0.292_{-0.100}^{+0.148}$ & $0.63_{-1.30}^{+4.20}$ & 32.0 / 70 & A \\
 & & & $8847.487_{-6.421}^{+5.514}$ & $0.049_{-0.041}^{+0.040}$ & $232.25_{-56.90}^{+104.18}$ & $-0.274_{-0.068}^{+0.056}$ & $0.325_{-0.120}^{+0.134}$ & $1.23_{-1.84}^{+9.43}$ & 34.0 / 70 & B \\
Gaia19dur & 283.8717 & -10.7520 & $8690.619_{-0.010}^{+0.012}$ & $-0.012_{-0.001}^{+0.001}$ & $130.78_{-5.30}^{+4.66}$ & $0.128_{-0.020}^{+0.015}$ & $-0.211_{-0.018}^{+0.019}$ & $-0.35_{-0.03}^{+0.03}$ & 221.0 / 179 & A \\
Gaia19duw & 298.1994 & 42.0715 & $8752.670_{-0.606}^{+1.418}$ & $0.148_{-0.026}^{+0.079}$ & $97.44_{-18.89}^{+9.00}$ & $-0.396_{-0.097}^{+0.034}$ & $-0.034_{-0.016}^{+0.006}$ & $0.69_{-0.57}^{+0.29}$ & 49.9 / 139 & A \\
Gaia19eca & 272.2020 & -24.8825 & $8719.303_{-0.003}^{+0.003}$ & $0.057_{-0.001}^{+0.001}$ & $83.43_{-0.15}^{+0.14}$ & $0.143_{-0.009}^{+0.008}$ & $0.286_{-0.001}^{+0.001}$ & $-0.07_{-0.01}^{+0.01}$ & 7865.1 / 1352 & A \\
Gaia19ect & 269.3059 & -29.4579 & $8755.940_{-0.006}^{+0.006}$ & $0.129_{-0.001}^{+0.001}$ & $84.24_{-0.67}^{+0.59}$ & $0.385_{-0.010}^{+0.009}$ & $-0.040_{-0.002}^{+0.002}$ & $-0.15_{-0.02}^{+0.02}$ & 13132.6 / 1323 & A \\
Gaia19eim & 200.7762 & -63.4019 & $8802.284_{-0.306}^{+0.257}$ & $-0.132_{-0.265}^{+0.327}$ & $63.53_{-20.86}^{+8.07}$ & $-0.040_{-0.109}^{+0.047}$ & $0.287_{-0.069}^{+0.200}$ & $0.05_{-0.65}^{+0.23}$ & 149.1 / 160 & A \\
Gaia19eku & 165.2140 & -61.7568 & $8796.102_{-3.405}^{+2.704}$ & $-0.211_{-0.498}^{+0.393}$ & $183.02_{-125.43}^{+274.51}$ & $0.003_{-0.227}^{+0.077}$ & $-0.158_{-0.038}^{+0.081}$ & $0.40_{-1.23}^{+1.57}$ & 51.4 / 113 & A \\
Gaia19enj & 303.0627 & 34.9701 & $8749.048_{-0.195}^{+2.618}$ & $-0.040_{-0.256}^{+0.092}$ & $73.63_{-20.48}^{+5.90}$ & $-0.370_{-0.177}^{+0.056}$ & $-0.181_{-0.073}^{+0.034}$ & $0.05_{-0.60}^{+0.12}$ & 401.8 / 257 & A \\
Gaia19esp & 194.2462 & -63.8131 & $8870.517_{-0.171}^{+0.191}$ & $-0.125_{-0.012}^{+0.015}$ & $98.85_{-7.60}^{+12.65}$ & $0.152_{-0.035}^{+0.047}$ & $-0.104_{-0.021}^{+0.026}$ & $0.67_{-0.16}^{+0.26}$ & 1085.0 / 404 & A \\
Gaia19evd & 158.1360 & -63.2424 & $8810.835_{-0.619}^{+0.836}$ & $-0.279_{-0.223}^{+0.424}$ & $96.10_{-33.17}^{+133.71}$ & $-0.078_{-0.139}^{+0.145}$ & $-0.189_{-0.067}^{+0.082}$ & $0.92_{-1.10}^{+4.07}$ & 81.3 / 150 & A \\
Gaia19fnb & 197.7973 & -65.3135 & $8857.737_{-1.345}^{+1.545}$ & $0.136_{-0.037}^{+0.060}$ & $108.97_{-24.93}^{+28.02}$ & $-0.196_{-0.098}^{+0.076}$ & $-0.198_{-0.066}^{+0.043}$ & $2.03_{-1.04}^{+1.21}$ & 142.7 / 172 & A \\
\bottomrule
\end{tabular*}
\tablefoot{Microlensing model parameters are reported as $x_{-\sigma_-}^{+\sigma_+}$, with uncertainties corresponding to the 16th and 84th percentiles.}
\end{table*}

\clearpage
\section{Lens parameters}
\begin{table*}[!htbp]
\caption{Lens mass and distance parameters for GSA alerts from 2017.}
\label{tab:lensMassDistGaia17}
\centering
\footnotesize
\setlength{\tabcolsep}{3pt}
\renewcommand{\arraystretch}{1.15}
\begin{tabular*}{\textwidth}{@{\extracolsep{\fill}}lccccccccc}
\toprule
Name & Model
& \multicolumn{4}{c}{$M_L \,\, [\rm{M}_\odot]$}
& \multicolumn{4}{c}{$D_L \, \, [\rm{kpc}]$} \\
\cmidrule(lr){3-6} \cmidrule(lr){7-10}
& 
& pyL+$D_S$ 
& pyL 
& BGM+$D_S$ 
& BGM 
& pyL+$D_S$ 
& pyL 
& BGM+$D_S$ 
& BGM \\
\midrule
Gaia17aqu & A & $0.46_{-0.07}^{+0.08}$ & $0.60_{-0.08}^{+0.10}$ & $0.42_{-0.21}^{+0.34}$ & $0.45_{-0.21}^{+0.33}$ & $2.84_{-0.47}^{+0.69}$ & $4.58_{-0.97}^{+1.87}$ & $2.84_{-0.70}^{+0.82}$ & $3.66_{-1.13}^{+1.59}$ \\
Gaia17bbs & A & $0.46_{-0.10}^{+0.12}$ & $0.50_{-0.09}^{+0.12}$ & $0.82_{-0.16}^{+0.21}$ & $0.77_{-0.25}^{+0.30}$ & $5.54_{-1.06}^{+1.31}$ & $6.13_{-1.19}^{+1.45}$ & $3.40_{-0.70}^{+0.92}$ & $2.74_{-0.63}^{+0.73}$ \\
Gaia17bdk & A & $0.33_{-0.07}^{+0.08}$ & $0.20_{-0.06}^{+0.10}$ & $0.98_{-0.17}^{+0.29}$ & $0.99_{-0.21}^{+0.24}$ & $9.41_{-0.68}^{+1.02}$ & $5.76_{-0.47}^{+0.67}$ & $5.93_{-1.29}^{+1.26}$ & $6.77_{-1.62}^{+2.02}$ \\
Gaia17bej & A & $0.61_{-0.06}^{+0.08}$ & $0.62_{-0.08}^{+0.10}$ & $0.50_{-0.20}^{+0.28}$ & $0.50_{-0.19}^{+0.28}$ & $3.39_{-0.62}^{+0.64}$ & $3.96_{-0.61}^{+0.87}$ & $3.60_{-1.05}^{+1.12}$ & $3.44_{-1.26}^{+1.50}$ \\
Gaia17cah & A & $0.10_{-0.01}^{+0.01}$ & $0.10_{-0.01}^{+0.02}$ & $0.10_{-0.03}^{+0.04}$ & $0.12_{-0.04}^{+0.04}$ & $4.42_{-0.33}^{+0.41}$ & $5.57_{-0.58}^{+0.47}$ & $4.28_{-0.48}^{+0.47}$ & $4.38_{-0.97}^{+1.09}$ \\
\bottomrule
\end{tabular*}
\tablefoot{Parameters are reported as $x_{-\sigma_-}^{+\sigma_+}$, with uncertainties corresponding to the 16th and 84th percentiles.}
\end{table*}

\begin{table*}[!htbp]
\caption{Lens mass and distance parameters for GSA alerts from 2018.}
\label{tab:lensMassDistGaia18}
\centering
\footnotesize
\setlength{\tabcolsep}{3pt}
\renewcommand{\arraystretch}{1.15}
\begin{tabular*}{\textwidth}{@{\extracolsep{\fill}}lccccccccc}
\toprule
Name & Model
& \multicolumn{4}{c}{$M_L \,\, [\rm{M}_\odot]$}
& \multicolumn{4}{c}{$D_L \, \, [\rm{kpc}]$} \\
\cmidrule(lr){3-6} \cmidrule(lr){7-10}
& 
& pyL+$D_S$ 
& pyL 
& BGM+$D_S$ 
& BGM 
& pyL+$D_S$ 
& pyL 
& BGM+$D_S$ 
& BGM \\
\midrule
Gaia18ajt & A & $0.11_{-0.01}^{+0.02}$ & $0.10_{-0.01}^{+0.01}$ & $0.12_{-0.03}^{+0.13}$ & $0.23_{-0.04}^{+0.05}$ & $4.27_{-0.47}^{+0.41}$ & $3.92_{-0.28}^{+0.35}$ & $3.67_{-1.20}^{+0.92}$ & $2.46_{-0.36}^{+0.74}$ \\
Gaia18akb & A & $0.35_{-0.02}^{+0.03}$ & $0.17_{-0.01}^{+0.01}$ & $0.11_{-0.01}^{+0.14}$ & $0.11_{-0.01}^{+0.01}$ & $0.51_{-0.04}^{+0.03}$ & $1.09_{-0.03}^{+0.06}$ & $1.28_{-0.61}^{+0.08}$ & $1.32_{-0.06}^{+0.07}$ \\
Gaia18bmt & A & $1.27_{-0.18}^{+0.24}$ & $0.85_{-0.10}^{+0.16}$ & $0.89_{-0.36}^{+0.32}$ & $0.77_{-0.26}^{+0.33}$ & $6.02_{-0.48}^{+0.44}$ & $2.78_{-0.29}^{+0.41}$ & $4.53_{-0.48}^{+0.53}$ & $5.03_{-2.18}^{+1.67}$ \\
 & B & $1.21_{-0.23}^{+0.21}$ & $1.43_{-0.16}^{+0.21}$ & $0.82_{-0.79}^{+0.40}$ & $0.85_{-0.34}^{+0.43}$ & $6.78_{-0.70}^{+0.76}$ & $8.61_{-1.65}^{+2.09}$ & $4.45_{-0.48}^{+1.03}$ & $5.06_{-1.74}^{+1.55}$ \\
Gaia18ceb & A & $0.10_{-0.01}^{+0.01}$ & $0.10_{-0.01}^{+0.01}$ & $0.74_{-0.11}^{+0.43}$ & $0.74_{-0.22}^{+0.24}$ & $4.50_{-0.63}^{+0.60}$ & $5.66_{-0.57}^{+0.68}$ & $2.94_{-0.58}^{+1.49}$ & $2.88_{-1.57}^{+1.64}$ \\
Gaia18cer & A & $0.09_{-0.01}^{+0.01}$ & $0.09_{-0.01}^{+0.01}$ & $0.07_{-0.01}^{+0.01}$ & $0.11_{-0.02}^{+0.01}$ & $3.21_{-0.24}^{+0.25}$ & $4.35_{-0.36}^{+0.30}$ & $3.16_{-0.24}^{+0.19}$ & $3.39_{-0.24}^{+0.31}$ \\
 & B & $0.10_{-0.01}^{+0.01}$ & $0.09_{-0.01}^{+0.01}$ & $0.25_{-0.17}^{+0.06}$ & $0.12_{-0.08}^{+0.02}$ & $3.66_{-0.24}^{+0.28}$ & $4.24_{-0.28}^{+0.27}$ & $1.69_{-0.20}^{+1.29}$ & $3.27_{-0.33}^{+0.74}$ \\
Gaia18ces & A & $0.09_{-0.01}^{+0.01}$ & $0.09_{-0.01}^{+0.01}$ & $0.20_{-0.03}^{+0.02}$ & $0.19_{-0.08}^{+0.03}$ & $2.28_{-0.21}^{+0.19}$ & $1.86_{-0.11}^{+0.13}$ & $1.45_{-0.10}^{+0.15}$ & $1.47_{-0.11}^{+1.17}$ \\
Gaia18cfc & A & $0.12_{-0.03}^{+0.03}$ & $0.13_{-0.02}^{+0.03}$ & $0.18_{-0.06}^{+0.14}$ & $0.16_{-0.05}^{+0.15}$ & $5.54_{-0.65}^{+0.72}$ & $6.32_{-0.71}^{+0.78}$ & $5.36_{-1.68}^{+1.14}$ & $3.97_{-0.84}^{+1.39}$ \\
Gaia18cfs & A & $0.10_{-0.01}^{+0.01}$ & $0.10_{-0.01}^{+0.01}$ & $0.77_{-0.13}^{+0.16}$ & $0.71_{-0.12}^{+0.15}$ & $9.63_{-0.72}^{+0.82}$ & $10.36_{-0.86}^{+0.61}$ & $3.32_{-0.57}^{+0.84}$ & $3.32_{-0.51}^{+0.71}$ \\
Gaia18cgp & A & $0.63_{-0.06}^{+0.09}$ & $0.47_{-0.07}^{+0.06}$ & $0.46_{-0.08}^{+0.10}$ & $0.41_{-0.09}^{+0.40}$ & $5.55_{-0.59}^{+0.59}$ & $2.40_{-0.26}^{+0.35}$ & $1.07_{-0.13}^{+0.12}$ & $1.62_{-0.74}^{+2.48}$ \\
Gaia18cib & A & $0.27_{-0.08}^{+0.12}$ & $0.33_{-0.09}^{+0.11}$ & $0.33_{-0.17}^{+0.31}$ & $0.37_{-0.23}^{+0.35}$ & $2.34_{-0.72}^{+1.08}$ & $3.53_{-1.41}^{+2.36}$ & $3.45_{-1.37}^{+1.37}$ & $3.84_{-1.73}^{+2.56}$ \\
Gaia18cik & A & $0.13_{-0.03}^{+0.05}$ & $0.14_{-0.03}^{+0.05}$ & $0.87_{-0.20}^{+0.25}$ & $0.81_{-0.21}^{+0.28}$ & $7.04_{-1.18}^{+1.36}$ & $7.37_{-1.44}^{+1.46}$ & $2.75_{-0.76}^{+1.21}$ & $2.16_{-0.61}^{+1.05}$ \\
 & B & $0.69_{-0.11}^{+0.11}$ & $0.69_{-0.09}^{+0.12}$ & $1.00_{-0.17}^{+0.24}$ & $1.01_{-0.19}^{+0.29}$ & $7.17_{-1.44}^{+1.40}$ & $7.41_{-1.25}^{+1.39}$ & $5.33_{-1.03}^{+1.34}$ & $4.65_{-1.53}^{+1.64}$ \\
Gaia18cjk & A & $0.83_{-0.10}^{+0.10}$ & $0.72_{-0.05}^{+0.06}$ & $0.29_{-0.19}^{+0.26}$ & $0.31_{-0.18}^{+0.34}$ & $3.32_{-0.46}^{+0.38}$ & $4.61_{-0.62}^{+0.98}$ & $1.84_{-0.75}^{+0.65}$ & $2.83_{-1.15}^{+1.81}$ \\
Gaia18clg & A & $0.26_{-0.05}^{+0.06}$ & $0.29_{-0.07}^{+0.05}$ & $0.40_{-0.20}^{+0.37}$ & $0.37_{-0.18}^{+0.31}$ & $2.27_{-0.26}^{+0.30}$ & $3.30_{-0.60}^{+0.86}$ & $1.82_{-0.57}^{+0.61}$ & $2.62_{-0.99}^{+1.27}$ \\
 & B & $0.21_{-0.07}^{+0.07}$ & $0.25_{-0.05}^{+0.07}$ & $0.32_{-0.15}^{+0.34}$ & $0.31_{-0.14}^{+0.30}$ & $2.25_{-0.35}^{+0.38}$ & $3.27_{-0.60}^{+0.88}$ & $1.70_{-0.57}^{+0.57}$ & $2.39_{-0.95}^{+1.16}$ \\
Gaia18cmk & A & $0.21_{-0.04}^{+0.04}$ & $0.25_{-0.05}^{+0.06}$ & $0.13_{-0.06}^{+0.09}$ & $0.12_{-0.06}^{+0.11}$ & $0.83_{-0.12}^{+0.13}$ & $1.01_{-0.15}^{+0.24}$ & $1.27_{-0.38}^{+0.49}$ & $1.61_{-0.65}^{+1.13}$ \\
Gaia18cnm & A & $0.10_{-0.01}^{+0.01}$ & $0.09_{-0.01}^{+0.01}$ & $0.45_{-0.10}^{+0.21}$ & $0.46_{-0.11}^{+0.26}$ & $7.53_{-1.06}^{+1.24}$ & $4.93_{-0.60}^{+0.82}$ & $1.45_{-0.40}^{+0.88}$ & $1.45_{-0.45}^{+0.83}$ \\
Gaia18cov & A & $0.53_{-0.06}^{+0.07}$ & $0.83_{-0.08}^{+0.08}$ & $0.33_{-0.15}^{+0.28}$ & $0.32_{-0.15}^{+0.27}$ & $1.72_{-0.21}^{+0.29}$ & $5.32_{-0.92}^{+0.68}$ & $1.80_{-0.50}^{+0.53}$ & $2.39_{-0.82}^{+1.23}$ \\
Gaia18cwh & A & $0.69_{-0.09}^{+0.07}$ & $0.79_{-0.11}^{+0.11}$ & $0.91_{-0.16}^{+0.21}$ & $0.90_{-0.17}^{+0.21}$ & $6.85_{-1.70}^{+1.02}$ & $8.50_{-1.48}^{+1.59}$ & $5.15_{-1.45}^{+2.12}$ & $5.14_{-1.59}^{+2.34}$ \\
Gaia18dep & A & $1.53_{-0.41}^{+0.12}$ & $0.53_{-0.04}^{+0.07}$ & $0.18_{-0.06}^{+0.51}$ & $0.18_{-0.09}^{+0.19}$ & $3.90_{-0.03}^{+0.04}$ & $0.70_{-0.09}^{+0.09}$ & $1.09_{-0.70}^{+0.24}$ & $1.37_{-0.55}^{+1.04}$ \\
Gaia18dif & A & $0.76_{-0.10}^{+0.09}$ & $0.87_{-0.12}^{+0.12}$ & $0.98_{-0.39}^{+0.43}$ & $0.87_{-0.35}^{+0.48}$ & $3.89_{-0.25}^{+0.32}$ & $6.79_{-0.76}^{+0.67}$ & $3.11_{-0.41}^{+0.53}$ & $4.44_{-1.21}^{+1.39}$ \\
 & B & $0.67_{-0.10}^{+0.10}$ & $0.79_{-0.11}^{+0.08}$ & $1.01_{-0.49}^{+0.49}$ & $0.79_{-0.34}^{+0.49}$ & $4.12_{-0.33}^{+0.45}$ & $7.61_{-0.68}^{+0.69}$ & $3.44_{-0.48}^{+0.56}$ & $4.55_{-1.16}^{+1.74}$ \\
Gaia18dma & A & $0.18_{-0.03}^{+0.06}$ & $0.14_{-0.01}^{+0.01}$ & $0.08_{-0.04}^{+0.07}$ & $0.09_{-0.05}^{+0.09}$ & $0.96_{-0.22}^{+0.12}$ & $1.25_{-0.13}^{+0.14}$ & $1.81_{-0.69}^{+1.09}$ & $1.60_{-0.67}^{+1.10}$ \\
 & B & $0.15_{-0.01}^{+0.02}$ & $0.17_{-0.02}^{+0.03}$ & $0.06_{-0.03}^{+0.06}$ & $0.07_{-0.03}^{+0.10}$ & $1.01_{-0.11}^{+0.13}$ & $0.88_{-0.12}^{+0.17}$ & $1.97_{-0.76}^{+0.88}$ & $1.92_{-0.84}^{+0.92}$ \\
Gaia18dml & A & $0.80_{-0.12}^{+0.18}$ & $0.91_{-0.16}^{+0.13}$ & $0.43_{-0.25}^{+0.43}$ & $0.47_{-0.26}^{+0.41}$ & $4.84_{-0.65}^{+0.85}$ & $6.34_{-1.02}^{+1.25}$ & $2.66_{-0.82}^{+0.96}$ & $4.86_{-1.88}^{+2.61}$ \\
Gaia18eaz & A & $0.09_{-0.01}^{+0.01}$ & $0.10_{-0.01}^{+0.02}$ & $0.23_{-0.10}^{+0.21}$ & $0.22_{-0.11}^{+0.22}$ & $3.15_{-0.42}^{+0.40}$ & $8.05_{-1.57}^{+1.85}$ & $2.27_{-0.67}^{+0.77}$ & $2.93_{-1.08}^{+1.58}$ \\
\bottomrule
\end{tabular*}
\tablefoot{Parameters are reported as $x_{-\sigma_-}^{+\sigma_+}$, with uncertainties corresponding to the 16th and 84th percentiles.}
\end{table*}

\begin{table*}[!htbp]
\caption{Lens mass and distance parameters for GSA alerts from 2019.}
\label{tab:lensMassDistGaia19}
\centering
\footnotesize
\setlength{\tabcolsep}{3pt}
\renewcommand{\arraystretch}{1.15}
\begin{tabular*}{\textwidth}{@{\extracolsep{\fill}}lccccccccc}
\toprule
Name & Model
& \multicolumn{4}{c}{$M_L \,\, [\rm{M}_\odot]$}
& \multicolumn{4}{c}{$D_L \, \, [\rm{kpc}]$} \\
\cmidrule(lr){3-6} \cmidrule(lr){7-10}
& 
& pyL+$D_S$ 
& pyL 
& BGM+$D_S$ 
& BGM 
& pyL+$D_S$ 
& pyL 
& BGM+$D_S$ 
& BGM \\
\midrule
Gaia19aqf & A & $0.59_{-0.06}^{+0.06}$ & $0.56_{-0.06}^{+0.12}$ & $0.35_{-0.17}^{+0.38}$ & $0.28_{-0.14}^{+0.28}$ & $2.43_{-0.30}^{+0.38}$ & $2.44_{-0.33}^{+0.54}$ & $2.56_{-0.91}^{+0.89}$ & $3.32_{-1.32}^{+1.87}$ \\
 & B & $0.22_{-0.05}^{+0.05}$ & $0.28_{-0.06}^{+0.07}$ & $0.25_{-0.13}^{+0.25}$ & $0.23_{-0.13}^{+0.23}$ & $2.23_{-0.34}^{+0.43}$ & $2.18_{-0.48}^{+0.65}$ & $2.28_{-1.03}^{+0.82}$ & $2.70_{-1.07}^{+1.89}$ \\
Gaia19aqw & A & $0.11_{-0.02}^{+0.02}$ & $0.11_{-0.01}^{+0.02}$ & $0.24_{-0.09}^{+0.24}$ & $0.31_{-0.14}^{+0.26}$ & $4.58_{-0.46}^{+0.54}$ & $4.63_{-0.50}^{+0.48}$ & $2.38_{-0.75}^{+1.02}$ & $1.69_{-0.62}^{+1.17}$ \\
Gaia19asx & A & $0.15_{-0.01}^{+0.01}$ & $0.19_{-0.01}^{+0.01}$ & $0.06_{-0.03}^{+0.07}$ & $0.06_{-0.04}^{+0.05}$ & $0.74_{-0.05}^{+0.05}$ & $0.59_{-0.02}^{+0.04}$ & $1.35_{-0.55}^{+0.66}$ & $1.40_{-0.49}^{+1.46}$ \\
 & B & $0.25_{-0.03}^{+0.02}$ & $0.16_{-0.02}^{+0.02}$ & $0.07_{-0.04}^{+0.08}$ & $0.07_{-0.03}^{+0.06}$ & $0.55_{-0.05}^{+0.08}$ & $0.82_{-0.08}^{+0.12}$ & $1.49_{-0.65}^{+0.85}$ & $1.44_{-0.55}^{+0.83}$ \\
Gaia19awc & A & $0.21_{-0.06}^{+0.09}$ & $0.34_{-0.10}^{+0.07}$ & $0.75_{-0.21}^{+0.22}$ & $0.64_{-0.25}^{+0.28}$ & $4.90_{-0.76}^{+0.62}$ & $9.72_{-1.32}^{+1.07}$ & $3.54_{-0.46}^{+0.62}$ & $3.40_{-1.62}^{+3.13}$ \\
Gaia19ayo & A & $0.33_{-0.05}^{+0.06}$ & $0.74_{-0.05}^{+0.05}$ & $0.29_{-0.15}^{+0.39}$ & $0.29_{-0.13}^{+0.28}$ & $3.23_{-0.45}^{+0.45}$ & $6.92_{-0.59}^{+0.97}$ & $4.38_{-1.05}^{+1.71}$ & $4.18_{-1.21}^{+2.29}$ \\
 & B & $0.40_{-0.07}^{+0.05}$ & $0.38_{-0.06}^{+0.07}$ & $0.23_{-0.12}^{+0.30}$ & $0.24_{-0.15}^{+0.29}$ & $2.92_{-0.55}^{+0.60}$ & $3.18_{-0.71}^{+0.59}$ & $4.16_{-1.12}^{+1.72}$ & $4.38_{-1.24}^{+3.01}$ \\
Gaia19bcm & A & $0.64_{-0.10}^{+0.13}$ & $0.60_{-0.10}^{+0.11}$ & $0.16_{-0.05}^{+0.12}$ & $0.24_{-0.12}^{+0.29}$ & $1.26_{-0.30}^{+0.63}$ & $1.28_{-0.30}^{+0.30}$ & $2.04_{-0.50}^{+0.70}$ & $1.39_{-0.63}^{+0.92}$ \\
Gaia19bht & A & $0.59_{-0.06}^{+0.06}$ & $0.58_{-0.06}^{+0.05}$ & $0.17_{-0.06}^{+0.30}$ & $0.18_{-0.06}^{+0.30}$ & $1.96_{-0.16}^{+0.15}$ & $2.38_{-0.20}^{+0.16}$ & $4.72_{-2.16}^{+0.51}$ & $4.68_{-2.13}^{+0.65}$ \\
Gaia19bhw & A & $0.80_{-0.12}^{+0.11}$ & $0.62_{-0.10}^{+0.14}$ & $0.39_{-0.14}^{+0.17}$ & $0.38_{-0.19}^{+0.20}$ & $7.41_{-1.13}^{+0.76}$ & $5.27_{-1.51}^{+1.99}$ & $3.69_{-1.02}^{+1.70}$ & $3.67_{-1.05}^{+2.95}$ \\
Gaia19bii & A & $0.58_{-0.06}^{+0.08}$ & $0.56_{-0.05}^{+0.06}$ & $0.10_{-0.03}^{+0.09}$ & $0.23_{-0.09}^{+0.09}$ & $1.97_{-0.21}^{+0.15}$ & $2.02_{-0.15}^{+0.23}$ & $5.18_{-1.22}^{+0.85}$ & $3.21_{-0.61}^{+0.92}$ \\
Gaia19bir & A & $0.41_{-0.12}^{+0.15}$ & $0.49_{-0.16}^{+0.14}$ & $0.36_{-0.21}^{+0.39}$ & $0.39_{-0.22}^{+0.37}$ & $3.89_{-1.47}^{+2.46}$ & $5.94_{-2.59}^{+3.05}$ & $3.65_{-1.25}^{+1.65}$ & $4.12_{-1.61}^{+2.22}$ \\
Gaia19bjq & A & $0.39_{-0.07}^{+0.08}$ & $0.38_{-0.06}^{+0.07}$ & $0.34_{-0.15}^{+0.31}$ & $0.34_{-0.14}^{+0.28}$ & $2.46_{-0.34}^{+0.40}$ & $2.71_{-0.32}^{+0.39}$ & $2.05_{-0.69}^{+0.89}$ & $2.27_{-0.85}^{+1.07}$ \\
Gaia19brt & A & $0.81_{-0.09}^{+0.15}$ & $0.99_{-0.11}^{+0.14}$ & $0.78_{-0.30}^{+0.50}$ & $0.60_{-0.26}^{+0.32}$ & $4.42_{-0.43}^{+0.40}$ & $5.96_{-0.74}^{+0.96}$ & $4.11_{-0.79}^{+0.63}$ & $5.89_{-1.58}^{+2.11}$ \\
Gaia19chi & A & $0.63_{-0.06}^{+0.08}$ & $0.75_{-0.08}^{+0.11}$ & $0.56_{-0.27}^{+0.33}$ & $0.50_{-0.21}^{+0.28}$ & $1.57_{-0.13}^{+0.20}$ & $4.37_{-0.49}^{+0.90}$ & $1.60_{-0.28}^{+0.31}$ & $3.07_{-0.88}^{+1.23}$ \\
 & B & $0.54_{-0.06}^{+0.09}$ & $0.62_{-0.08}^{+0.08}$ & $0.35_{-0.17}^{+0.33}$ & $0.34_{-0.16}^{+0.28}$ & $1.18_{-0.16}^{+0.25}$ & $2.61_{-0.57}^{+0.95}$ & $1.36_{-0.33}^{+0.35}$ & $2.32_{-0.80}^{+1.05}$ \\
Gaia19cyv & A & $0.08_{-0.01}^{+0.00}$ & $0.08_{-0.01}^{+0.01}$ & $0.09_{-0.03}^{+0.15}$ & $0.08_{-0.05}^{+0.09}$ & $2.21_{-0.12}^{+0.23}$ & $2.60_{-0.18}^{+0.17}$ & $1.62_{-0.70}^{+0.46}$ & $1.88_{-0.76}^{+0.76}$ \\
 & B & $0.08_{-0.01}^{+0.00}$ & $0.08_{-0.00}^{+0.01}$ & $0.14_{-0.07}^{+0.18}$ & $0.11_{-0.05}^{+0.09}$ & $2.53_{-0.16}^{+0.19}$ & $3.68_{-0.25}^{+0.19}$ & $1.70_{-0.57}^{+0.67}$ & $2.01_{-0.72}^{+0.72}$ \\
Gaia19dae & A & $0.67_{-0.08}^{+0.12}$ & $0.71_{-0.08}^{+0.10}$ & $0.61_{-0.26}^{+0.30}$ & $0.45_{-0.20}^{+0.31}$ & $1.63_{-0.17}^{+0.22}$ & $2.49_{-0.28}^{+0.42}$ & $1.66_{-0.31}^{+0.35}$ & $2.54_{-0.98}^{+1.18}$ \\
 & B & $0.88_{-0.13}^{+0.15}$ & $0.79_{-0.10}^{+0.08}$ & $0.83_{-0.32}^{+0.45}$ & $0.67_{-0.26}^{+0.29}$ & $3.23_{-0.28}^{+0.30}$ & $3.08_{-0.28}^{+0.17}$ & $1.84_{-0.28}^{+0.30}$ & $3.25_{-1.24}^{+1.30}$ \\
Gaia19dak & A & $0.09_{-0.01}^{+0.01}$ & $0.09_{-0.01}^{+0.01}$ & $0.27_{-0.17}^{+0.07}$ & $0.27_{-0.14}^{+0.06}$ & $2.72_{-0.15}^{+0.20}$ & $4.03_{-0.28}^{+0.32}$ & $1.47_{-0.26}^{+0.99}$ & $1.40_{-0.33}^{+0.49}$ \\
 & B & $0.18_{-0.03}^{+0.02}$ & $0.25_{-0.06}^{+0.05}$ & $0.28_{-0.07}^{+0.09}$ & $0.31_{-0.08}^{+0.15}$ & $2.43_{-0.21}^{+0.27}$ & $2.37_{-0.34}^{+0.53}$ & $1.72_{-0.28}^{+0.66}$ & $1.58_{-0.45}^{+0.63}$ \\
Gaia19drc & A & $0.39_{-0.04}^{+0.05}$ & $0.41_{-0.04}^{+0.04}$ & $0.37_{-0.02}^{+0.02}$ & $0.37_{-0.03}^{+0.02}$ & $1.28_{-0.11}^{+0.10}$ & $1.47_{-0.13}^{+0.13}$ & $1.43_{-0.09}^{+0.11}$ & $1.04_{-0.06}^{+0.28}$ \\
Gaia19dtd & A & $0.19_{-0.07}^{+0.12}$ & $0.21_{-0.06}^{+0.09}$ & $0.55_{-0.19}^{+0.31}$ & $0.51_{-0.26}^{+0.31}$ & $6.21_{-0.74}^{+1.03}$ & $9.14_{-1.62}^{+1.01}$ & $4.90_{-1.32}^{+1.30}$ & $5.06_{-1.66}^{+1.30}$ \\
 & B & $0.15_{-0.04}^{+0.06}$ & $0.17_{-0.06}^{+0.07}$ & $0.74_{-0.22}^{+0.31}$ & $0.80_{-0.29}^{+0.30}$ & $9.68_{-0.54}^{+0.46}$ & $11.21_{-0.96}^{+0.70}$ & $5.57_{-0.72}^{+0.83}$ & $5.59_{-1.20}^{+1.50}$ \\
Gaia19dtk & A & $0.36_{-0.07}^{+0.10}$ & $0.40_{-0.09}^{+0.14}$ & $0.31_{-0.15}^{+0.29}$ & $0.31_{-0.16}^{+0.29}$ & $1.27_{-0.36}^{+0.46}$ & $1.52_{-0.45}^{+0.64}$ & $1.88_{-0.74}^{+1.19}$ & $1.86_{-0.74}^{+1.64}$ \\
 & B & $0.49_{-0.09}^{+0.10}$ & $0.60_{-0.11}^{+0.14}$ & $0.27_{-0.14}^{+0.29}$ & $0.26_{-0.14}^{+0.27}$ & $1.75_{-0.48}^{+0.66}$ & $2.62_{-0.60}^{+2.07}$ & $1.90_{-0.67}^{+1.07}$ & $1.90_{-0.68}^{+1.32}$ \\
Gaia19dur & A & $0.10_{-0.01}^{+0.02}$ & $0.10_{-0.01}^{+0.02}$ & $0.39_{-0.18}^{+0.31}$ & $0.44_{-0.19}^{+0.31}$ & $4.59_{-0.41}^{+0.40}$ & $6.56_{-0.93}^{+0.74}$ & $2.79_{-0.76}^{+1.07}$ & $2.75_{-0.97}^{+1.36}$ \\
Gaia19duw & A & $0.55_{-0.07}^{+0.06}$ & $0.51_{-0.09}^{+0.09}$ & $0.28_{-0.12}^{+0.21}$ & $0.29_{-0.13}^{+0.23}$ & $1.29_{-0.29}^{+0.41}$ & $2.05_{-0.56}^{+0.93}$ & $2.04_{-0.65}^{+1.00}$ & $1.82_{-0.64}^{+1.03}$ \\
Gaia19eca & A & $0.13_{-0.02}^{+0.04}$ & $0.18_{-0.02}^{+0.03}$ & $0.10_{-0.03}^{+0.03}$ & $0.05_{-0.02}^{+0.03}$ & $4.21_{-0.59}^{+0.33}$ & $4.03_{-0.34}^{+0.36}$ & $3.75_{-0.34}^{+0.39}$ & $5.92_{-0.80}^{+0.98}$ \\
Gaia19ect & A & $0.09_{-0.01}^{+0.01}$ & $0.09_{-0.01}^{+0.01}$ & $0.04_{-0.01}^{+0.39}$ & $0.10_{-0.04}^{+0.04}$ & $5.14_{-0.34}^{+0.29}$ & $5.51_{-0.41}^{+0.39}$ & $6.63_{-1.74}^{+0.84}$ & $3.39_{-1.11}^{+3.55}$ \\
Gaia19eim & A & $0.36_{-0.08}^{+0.09}$ & $0.51_{-0.11}^{+0.12}$ & $0.54_{-0.23}^{+0.41}$ & $0.72_{-0.26}^{+0.29}$ & $1.61_{-0.41}^{+0.52}$ & $4.24_{-1.51}^{+1.91}$ & $1.48_{-0.36}^{+0.34}$ & $3.63_{-1.49}^{+1.82}$ \\
Gaia19eku & A & $0.59_{-0.09}^{+0.12}$ & $0.62_{-0.09}^{+0.11}$ & $0.42_{-0.24}^{+0.42}$ & $0.43_{-0.24}^{+0.43}$ & $4.79_{-1.13}^{+1.56}$ & $5.68_{-1.57}^{+2.91}$ & $4.63_{-1.34}^{+1.66}$ & $4.72_{-1.57}^{+2.25}$ \\
Gaia19enj & A & $0.24_{-0.08}^{+0.08}$ & $0.27_{-0.08}^{+0.09}$ & $0.25_{-0.13}^{+0.25}$ & $0.25_{-0.12}^{+0.27}$ & $2.56_{-0.72}^{+0.77}$ & $3.38_{-1.12}^{+1.50}$ & $2.71_{-0.92}^{+0.99}$ & $2.41_{-0.90}^{+1.47}$ \\
Gaia19esp & A & $0.76_{-0.08}^{+0.10}$ & $0.75_{-0.10}^{+0.17}$ & $0.71_{-0.15}^{+0.20}$ & $0.67_{-0.16}^{+0.23}$ & $3.40_{-0.50}^{+0.59}$ & $4.44_{-0.66}^{+0.85}$ & $2.92_{-0.51}^{+0.79}$ & $2.34_{-0.58}^{+0.69}$ \\
Gaia19evd & A & $0.66_{-0.11}^{+0.11}$ & $0.75_{-0.10}^{+0.15}$ & $0.37_{-0.21}^{+0.40}$ & $0.45_{-0.25}^{+0.42}$ & $3.30_{-0.76}^{+0.69}$ & $5.45_{-1.33}^{+1.67}$ & $2.68_{-0.73}^{+0.66}$ & $5.05_{-2.00}^{+3.25}$ \\
Gaia19fnb & A & $0.58_{-0.09}^{+0.14}$ & $0.72_{-0.11}^{+0.13}$ & $0.28_{-0.15}^{+0.32}$ & $0.29_{-0.16}^{+0.32}$ & $2.44_{-0.73}^{+0.97}$ & $6.25_{-1.35}^{+1.42}$ & $2.37_{-0.77}^{+0.91}$ & $3.61_{-1.41}^{+1.90}$ \\
\bottomrule
\end{tabular*}
\tablefoot{Parameters are reported as $x_{-\sigma_-}^{+\sigma_+}$, with uncertainties corresponding to the 16th and 84th percentiles.}
\end{table*}

\clearpage
\section{BHTOM observations}

\begin{table*}[!htbp]
\caption{Events observed by each instrument used for follow-up of the GSA microlensing events.}
\label{tab:telescopeInfo}
\centering
\footnotesize
\begin{tabularx}{\textwidth}{@{}l >{\raggedright\arraybackslash}X@{}}
\hline\hline
\textbf{Telescope Name/Code} & \textbf{Events Observed} \\
\hline
TRT SBO 70-cm & Gaia18dif, Gaia19bcm \\
TRT TNO 50-cm & Gaia17aqu \\
PROMPT8 60-cm & Gaia18bmt, Gaia18cjk, Gaia18dep \\
LCO coj 100-cm & Gaia17aqu, Gaia17bbs, Gaia17bdk, Gaia18bmt, Gaia18dif, Gaia18eaz, Gaia18cjk, Gaia18dep, Gaia19esp \\
LCO coj 200-cm & Gaia18bmt, Gaia19eim, Gaia19evd, Gaia19fnb \\
LCO cpt 100-cm & Gaia17aqu, Gaia17bbs, Gaia17bdk, Gaia18bmt, Gaia18dif, Gaia18eaz, Gaia18cjk, Gaia19esp \\
LCO lsc 100-cm & Gaia17aqu, Gaia17bbs, Gaia17bdk, Gaia18bmt, Gaia18dif, Gaia18eaz, Gaia18cjk, Gaia18dep, Gaia19esp \\
LCO ogg 200-cm & Gaia19dak, Gaia19evd \\
SMARTS 130-cm & Gaia17bej, Gaia18ajt, Gaia18bmt, Gaia18clg, Gaia18cov, Gaia18dep, Gaia18dif, Gaia18eaz, Gaia19aqw \\
Rapid Eye Mount 60-cm & Gaia18bmt, Gaia18cjk, Gaia18dep, Gaia18eaz, Gaia19aqw \\
TÜBİTAK T 100-cm & Gaia18cjk \\
Ondrejov 50-cm & Gaia18cjk, Gaia18cmk \\
Joan Oró Telescope 80-cm & Gaia18cjk, Gaia18cmk \\
Mol\.{e}tai 165-cm & Gaia18cmk \\
C2PU 104-cm & Gaia18cmk \\
pt5m 50-cm & Gaia18cmk, Gaia19dak \\
Aristarchos 230-cm & Gaia19cyv, Gaia19dak \\
Loiano 152-cm & Gaia19cyv, Gaia19dak \\
\hline
\end{tabularx}
\end{table*}

\begin{table*}[!htbp]
\caption{Instruments used for ground-based follow-up of the GSA microlensing events.}
\label{tab:telescopeInfo}
\centering
\footnotesize
\begin{tabularx}{\textwidth}{
    l
    >{\hsize=9\hsize\arraybackslash}X
    >{\hsize=6\hsize\arraybackslash}X
    l l l l
}
\hline\hline
\textbf{Telescope Name/Code} & \textbf{Observatory Name} & \textbf{Location} & \textbf{Long. [deg]} & \textbf{Lat. [deg]} & \textbf{Alt. [m]} & \textbf{Reference} \\
\hline
TRT SBO 70-cm & Springbrook Observatory & New South Wales, Australia & 149.1 & -31.3 & 1020 & - \\
TRT TNO 50-cm & Thai National Observatory & Chiang Mai, Thailand & 98.48 & 18.57 & 2450 & - \\
PROMPT8 60-cm & Cerro Tololo Inter-American Observatory & Coquimbo, Chile & -70.8 & -30.2 & 2286 & \citet{PromptObservatory} \\
LCO coj 100-cm & Siding Spring Observatory & Siding Spring, Australia & 149.1 & -31.3 & 1165 & \citet{LCOBrown} \\
LCO coj 200-cm & Siding Spring Observatory & Siding Spring, Australia & 149.1 & -31.3 & 1165 & \citet{LCOBrown} \\
LCO cpt 100-cm & South African Astronomical Observatory & Sutherland, South Africa & 20.8 & -32.4 & 1760 & \citet{LCOBrown} \\
LCO lsc 100-cm & Cerro Tololo Inter-American Observatory & Coquimbo, Chile & -70.8 & -30.2 & 2201 & \citet{LCOBrown} \\
LCO ogg 200-cm & Haleakalā Observatory & Maui, USA & -156.2560 & 20.7075 & 3055 & \citet{LCOBrown} \\
SMARTS 130-cm & Cerro Tololo Inter-American Observatory & Coquimbo, Chile & -70.8 & -30.2 & 2215 & - \\
Rapid Eye Mount 60-cm & La Silla Observatory & Coquimbo, Chile & -70.7 & -29.3 & 2400 & \citet{REMobservatory} \\
TÜBİTAK T 100-cm & TÜBİTAK National Observatory & Antalya, Turkey & 30.3356 & 36.8242 & 2500 & - \\
Ondrejov 50-cm & Ondřejov Observatory & Ondřejov, Czech Republic & 14.7814 & 49.9094 & 527 & \citet{OndrejovObs} \\
Joan Oró Telescope 80-cm & The Montsec Observatory & Montsec, Spain & 0.7297 & 42.0517 & 1570 & \citet{JoanOroTelescope} \\
Mol\.{e}tai 165-cm & Mol\.{e}tai Astronomical Observatory & Kulionys, Lithuania & 25.5631 & 55.3159 & 200 & - \\
C2PU 104-cm & Observatoire de la Côte d'Azur & Calern, France & 6.9230 & 43.7537 & 1260 & - \\
pt5m 50-cm & Roque de los Muchachos Observatory & La Palma, Spain & -17.8792 & 28.7606 & 2326 & \citet{pt5mtelescope} \\
Aristarchos 230-cm & Helmos Observatory & Klitoria, Greece & 22.1983 & 37.9856 & 2340 & \citet{HelmosOBs} \\
Loiano 152-cm & Loiano Observatory & Bologna, Italy & 11.3344 & 44.2591 & 785 & - \\
\hline
\end{tabularx}
\end{table*}

\end{appendix}

\end{document}